\documentclass[useAMS,usenatbib]{mnras}

\usepackage{graphicx}
\usepackage{float}

\usepackage{bm}
\usepackage{amsmath}
\usepackage{amsfonts,amssymb}

\usepackage{color}
\usepackage{soul} 
\usepackage{enumerate}
\usepackage{float}

\usepackage{tabularx}
\usepackage{multicol}        
\usepackage{bm}	
\usepackage{booktabs}
\usepackage{array} 
\usepackage{diagbox} 

\usepackage{siunitx}
\usepackage{booktabs}  
\usepackage{comment}

\usepackage{hyperref}
\usepackage{tabularx}

\usepackage{pdfpages}  

 \title[Extended dark-energy signatures]{Signatures of Extended Dark Energy Parametrisations in Structure Formation under Background Constraints}

\author[Greco A. Pe\~na et al. ]{Greco A. Pe\~na$^{1,2}$\thanks{E-mail:greco.pena@postgrado.uv.cl}, Mario H. Amante$^{2}$, Javier Chagoya$^{2}$, Cristian Barrera-Hinojosa$^{1}$, \newauthor 
C. Ortiz$^{2}$ and Graeme Candlish$^{1}$\\
$^{1}$Instituto de F\'isica y Astronom\'ia, Facultad de Ciencias, \\Universidad de Valpara\'iso, Avda. Gran Breta\~na 1111, Valpara\'iso, Chile. \\
$^{2}$Unidad Acad\'emica de F\'isica, Universidad Aut\'onoma de Zacatecas, \\Calzada Solidaridad esquina con Paseo a la Bufa S/N C.P. 98060, Zacatecas, M\'exico.}

\date{Accepted YYYYMMDD. Received YYYYMMDD; in original form YYYYMMDD}
\pubyear{2025}

\begin{document}

\maketitle

\begin{abstract}

We study structure formation in alternative cosmological models constrained by background observations, including $\Lambda$CDM, $w$CDM, the Chevallier–Polarski–Linder parameterisation and a flexible Chebyshev expansion of the dark energy equation of state. The models are constrained using baryon acoustic oscillations, cosmic microwave background, cosmic chronometers and strong lensing measurements. Using the best-fitting parameters, we generate cosmology-dependent initial conditions and perform $N$-body simulations to analyse the matter power spectrum, halo mass function and halo density profiles. Although all models remain broadly consistent with $\Lambda$CDM at the background level, differences in the physical matter density $\Omega_{0m}h^2$ and in the expansion history $H(z)$ lead to distinct growth histories that are amplified by non-linear evolution. We find a clear hierarchy in the power spectrum amplitude and in $\sigma_8$, with the Chebyshev and CPL models exhibiting enhanced small-scale power, earlier halo formation at $z\gtrsim2$ and a migration of excess toward higher masses at late times. The $w$CDM model displays milder and partially compensating effects driven by its different expansion history. When expressed in terms of the scaled radius $r/R_{200\mathrm{c}}$, halo density profiles show a high degree of universality across cosmologies, indicating that internal halo structure is largely governed by the same gravitational dynamics. These results demonstrate that even modest background-level variations in $w(z)$ can translate into coherent non-linear signatures, highlighting the constraining power of large-scale structure observables in extended dark energy models.
\end{abstract}


\begin{keywords}
methods: numerical -- cosmology: theory -- cosmology: dark energy -- cosmology: observations -- cosmology: large-scale structure of Universe 
\end{keywords}

\section{Introduction}\label{sec:introduction}

The formation and evolution of large-scale structure (LSS) stand at the core of modern cosmology, as they encode the physical processes that shaped the present-day cosmic web of galaxies and clusters~\citep{Angulo:2022,Springel2006,Vogelsberger2020}. Tracing the distribution and growth of these structures provides powerful constraints on the fundamental properties of dark matter and dark energy, two essential yet still enigmatic components of the Universe \citep{Copeland2006,Caldwell2009}. Despite the success of the $\Lambda$CDM model in explaining a wide range of observations, particularly those from the early Universe such as measurements of the cosmic microwave background (CMB) by the Planck Collaboration \citep{Planck:2020}, increasing tensions have emerged over the past decade, most notably the discrepancy between local and CMB determinations of the Hubble constant \citep{Riess2019,Riess2022,Verde2019} and inconsistencies in the growth rate of structures inferred from weak-lensing surveys \citep{Heymans2021,Asgari2021,Abbott2022}. Moreover, the nature of dark energy remains uncertain: it could correspond to a cosmological constant, a dynamical field, or arise from new physical mechanisms beyond the standard paradigm \citep{Wang2018}. Recent observational probes, such as the first- and second-year (DR1 and DR2) DESI baryon acoustic oscillation (BAO) measurements \citep{DESI:2024mwx,DES:CPL:2025,karimDESI2025,william:2025,reeves:2025}, continue to suggest the possibility of time-evolving dark energy.

Alternative cosmological models, particularly those involving a dynamical dark energy component or modifications to the equation of state, provide viable frameworks to address these emerging tensions \citep{Copeland2006,Caldwell2009,Huterer2017,Menci:2024,lapi:2025,Akarsu:2025}. Such models can remain consistent with current background observations while leading to distinctive predictions for structure formation, especially in the non-linear regime \citep{Amendola2010,Wang2018,Oliveira2025}. It is therefore essential to investigate not only the expansion history of the Universe, but also how these scenarios impact the growth of structure across different scales.

Historically, analytic works have been made to explore dynamical dark energy models (e.g., \cite{Wang1998,Linder:2003}) and their  effects in the evolution of the Universe. While many studies focus on the linear growth of perturbations in dark energy models with either a constant equation-of-state parameter $w$ or a time-varying parameter $w(z)$ \citep{Basilakos2003, kulinich:2004,reyhani:2024,Menci:2024,lapi:2025, Akarsu:2025}, analytical descriptions of non-linear structure formation in these scenarios remain challenging. 
As a result, the standard approach to study the fully non-linear regime relies on $N$-body simulations. In this context, several works have investigated the impact of alternative models on structure formation using numerical simulations \citep{Linder:2005,Casarini:2009,Alimi:2010,Pfeifer:2020,Sokoliuk:2023,beltz-Mohrmann:2025,Ishiyama2025}.

In~\cite{beltz-Mohrmann:2025}, the impact of $\Lambda$CDM and Chevallier-Polarski-Linder CPL ($w_0w_1$) cosmologies on linear and non-linear scales was explored using cosmological parameters derived from DESI DR1 analysis~\citep{DESI:2024mwx}, which incorporates constraints from BAO and CMB priors. Similarly, \cite{Ishiyama2025} performed a comparison of the same cosmological scenarios employing a significantly larger simulation volume to mitigate cosmic variance, thereby enabling a more robust comparison with DESI observations.

In this work, we investigate four cosmological scenarios: the $\Lambda$CDM model, a constant $w$CDM model \citep{TurnerWhite1997}, the CPL parameterisation \citep{Chevallier:2000qy,Linder:2003} and a Chebyshev expansion model for the dark energy equation of state \citep{Zunckel2007,Ma2011}. 

We first constrain the background evolution of each model using current observational data, including BAO \citep{Eisenstein:1997ik,Alam2017}, CMB distance priors \citep{Komatsu:2011}, cosmic chronometers (CC) \citep{Jimenez:2001gg,Moresco2012} and strong lensing systems (SLS) \citep{Refsdal1964}. The resulting best-fitting cosmological parameters are then used to generate consistent initial conditions for the $N$-body simulations, allowing us to follow the non-linear evolution of matter and the formation of dark matter haloes in each scenario.


By jointly analysing the background evolution and the non-linear growth of structure in these models, we assess how deviations in the dark energy sector propagate into observable signatures in the matter distribution. In particular, we examine redshift and scale ranges where alternative cosmologies begin to depart from $\Lambda$CDM predictions. This approach provides a physically consistent framework to connect background constraints with non-linear clustering observables, highlighting distinctive signatures that may be tested with current and forthcoming surveys such as DESI, Euclid, the Vera C. Rubin Observatory and the Nancy Grace Roman Space Telescope~\citep{Laureijs2011,LSST2009,Spergel2015,DESI2016,novell:2025}.


The paper is organised as follows. In Section~\ref{CM}, we introduce the cosmological models under consideration and describe their parameterisation. Section~\ref{sec:data} presents the data sets and methodology employed to constrain the background evolution using BAO, CMB, CC and SLS, and describes how the resulting cosmological parameters are used to construct consistent initial conditions and implement the alternative cosmological models in the $N$-body simulations. Section~\ref{sec:Res} reports our results, divided into two parts: Subsection~\ref{subsec:backgound_const} presents the outcomes of a Bayesian statistical analysis based on a joint fit to the data, while Subsection~\ref{subsec:non-linear} investigates the implications of the constrained models for cosmic structure formation using $N$-body simulations. Finally, Section~\ref{sec:discussion} discusses these results and summarises our findings, highlighting the impact of the considered cosmological models on large-scale structure.

\section{Cosmological models} \label{CM}
In this section, we present the cosmological models that will be tested with the observational data described in Sec.~\ref{sec:data}. In all cases, we restrict to models with flat spatial geometry containing dark and baryonic matter, dark energy and radiation.
\subsection{\texorpdfstring{$\Lambda$}{L}CDM}


The $\Lambda$CDM model, regarded as the standard cosmological model, offers a coherent framework for understanding the evolution and structure of the Universe. It describes the background expansion history, composition and geometry in terms of the Hubble constant, $H_0$, and the dark matter, baryon, radiation and dark energy density parameters at the present epoch ($z=0$)\footnote{In the general case, one should consider the curvature density parameter as well; however, throughout this work we assume a spatially flat Friedmann--Robertson--Walker geometry.}: $\Omega_{0\textrm{CDM}}$, $\Omega_{0b}$, $\Omega_{0r}$ and $\Omega_\Lambda$, respectively. The matter density parameter is defined as $\Omega_{0m}\equiv\Omega_{0\textrm{CDM}} + \Omega_{0b}$, and dark energy is modelled by a cosmological constant $\Lambda$, which can be interpreted as a fluid with constant density and equation of state (EoS) $p_\Lambda=-\rho_\Lambda$, i.e. with $w=-1$.

Defining the dimensionless Hubble parameter $E(z) \equiv H(z)/H_0$, the Friedmann equation for the standard cosmological model is
\begin{align}\label{eq:lcdm}
E^2(z) _{\Lambda \textrm{CDM}} &=
\Omega_{0m}(1+z)^3
+ \Omega_{0r}(1+z)^4
+ (1 - \Omega_{0m} - \Omega_{0r}),
\end{align}
where we have used the Friedmann constraint to express the density parameter of the cosmological constant in terms of the matter and radiation density parameters. We adopt a unified approach for computing the radiation component across all cosmological models. The radiation density parameter is computed as 
$\Omega_{0r} = \Omega_{\gamma 0}\left[1 + 0.2271\, N_{\mathrm{eff}}\right]$, 
where $\Omega_{\gamma 0} = 2.469 \times 10^{-5} h^{-2}$ is the photon density parameter, $h$ denotes the reduced Hubble parameter, and $N_{\mathrm{eff}} = 3.046$ is the effective number of relativistic neutrino species \citep{Komatsu:2011}.

\subsection{$w$CDM}
This model is the simplest modification to $\Lambda$CDM. Here, dark energy has a constant EoS parameter but it deviates from $w_0=-1$. In order to obtain an accelerated Universe, it should satisfy $w_0<-1/3$. In this model, the equation $E(z)$ can be written as: 
\begin{eqnarray}
E(z)_{w}^2&=&\Omega_{0m}(1+z)^3  + \Omega_{0r}(1+z)^4 \nonumber \\
& &+(1 -\Omega_{0m}-\Omega_{0r})(1+z)^{3(1+w_0)}.
\end{eqnarray}
This model has one additional free parameter ($w_0$) with respect to $\Lambda$CDM.

\subsection{CPL}\label{CPL}

The Chevallier-Polarski-Linder (CPL)~\cite{Chevallier:2000qy,Linder:2003} 
parameterisation 
is an approach to study dynamical DE models through a parameterisation of its EoS given by 

\begin{equation}
w(z)=w_0+w_1\frac{z}{(1+z)},
\label{eq:eosCPL}
\end{equation}
where $w_{0}$ is the EoS parameter at redshift $z=0$ and $w_{1}=\mathrm{d}w/\mathrm{d}z|_{z=0}$. The dimensionless $E(z)$ for the CPL parameterisation is 
\begin{align}
E(z)_{\textrm{CPL}}^2&=\Omega_{m0}(1+z)^3  + \Omega_{0r}(1+z)^4 \nonumber\\& +(1-\Omega_{0m}-\Omega_{0r})(1+z)^{3(1+w_0+w_{1})}\exp\left(\frac{-3w_1z}{1+z}\right).\label{CPLE}
\end{align}
This model contains four free parameters, two more than $E(z)_{\Lambda \textrm{CDM}}$.

\subsection{Chebyshev} \label{Cheb}
A series truncated at fourth order using the Chebyshev polynomials was recently used in~\cite{DESI:2024aqx} for modelling the EoS parameter of dark energy as,
\begin{align}\label{eq:w1}
\tilde{w}_1(z) 
& = -\left( C_0 + C_1 x + C_2 (2x^2 -1 ) + C_3(4x^3 - 3x) \right),
\end{align}
where 
\begin{align}
x & = 1 - \frac{2(z_{\max}-z)}{z_{\max}-z_{\min}}.
\end{align}
Notice that $x\in[-1,1]$, with $x=-1$ when $z=z_{\min}$ and $x=1$ when $z=z_{\max}$. As stated previously in \cite{DESI:2024aqx} we choose $z_{\min}=0$ and $z_{\max}=3.5$.
In order to examine $w(z)$ using high-redshift data, such as the CMB, a parameterisation valid for $z>z_{\max}$ is required. Following the same approach we employ a transition function that ensures $w(z)=-1$ at early times.
The transition function is chosen as
\begin{align}\label{eq:w2}
 \tilde{w}_{2}(z) & = -1+ \left(B_0+B_1 u\right) \exp\left({-\frac{u^2}{\Delta ^2}}\right), \\ \nonumber
 u&=\ln \left(\frac{1+z}{1+z_{\max}}\right),
\end{align}
and is valid for $z\geq z_{\max}$. Assuming $z_{\min} = 0$, the matching conditions $\tilde{w}_1(z_{\max})=\tilde{w}_2(z_{\max}), \tilde{w}_1'(z_{\max})=\tilde{w}_2'(z_{\max})$ and $\tilde{w_1}''(z_{\max})=\tilde{w}_2''(z_{\max})$ lead to 
\begin{subequations}\label{eq:match}
\begin{align}
B_0 & =  -C_0-C_1-C_2-C_3+1, \\
B_1 & =  -\frac{2 \left(C_1+4 C_2+9 C_3\right) \left(z_{\max }+1\right)}{z_{\max }}, \\
\Delta^2 & =\frac{\left(1-C_0-C_1-C_2-C_3\right) \left(z_{\max }+1\right)^{-1}z_{\max }^2}{ C_1 z_{\max }+4 C_2 \left(3 z_{\max }+2\right)+C_3 \left(57 z_{\max }+48\right)}.
\end{align}
\end{subequations}

The full form of the EoS is \footnote{Note that both $\tilde{w}_1(z)$ and $\tilde{w}_2(z)$ are functions describing two regimes in the Chebyshev model, whereas $w_0$ and $w_1$ are the parameters of CPL parametrization.}
\begin{equation}
    w(z) = \left\{\begin{array}{ll} 
    \tilde{w}_1(z) & \text{ if } z\leq z_{\max}, \\
    \tilde{w}_2(z) & \text{ if } z\geq z_{\max}, \\
    \end{array} \right.
\end{equation}
with the parameters of $\tilde{w}_2(z)$ given in Eqs.~\eqref{eq:match}. The dimensionless Friedmann equation is constructed as
\begin{align}
    E(z)_{\textrm{Ch}}^2&=\Omega_{m0}(1+z)^3  + \Omega_{0r}(1+z)^4 +\Omega_{\rm DE}(z)\, ,
\end{align}
where
\begin{equation}
    \Omega_{\rm DE}(z) = \Omega_{0{\rm DE}}\exp\left[{3\int \frac{1+w({\tilde{z}})}{{1+\tilde{z}}}d{\tilde{z}}}\right]\, ,
\end{equation}
and $\Omega_{0m} + \Omega_{0r} + \Omega_{0{\rm DE}} = 1$. Here we have four additional free parameters ($C_0, C_1, C_2, C_3$) with respect to $E(z)_{\Lambda \textrm{CDM}}$.
\section{Data and Methodology} \label{sec:data}
The 
models discussed in Section~\ref{CM} are next constrained using observations from baryon acoustic oscillations (BAO), 
cosmic chronometers (CC), cosmic microwave background (CMB) radiation and strong lensing systems (SLS). In the following, we provide a brief introduction to these data sets, present the methodology for obtaining the cosmological parameters and finally, outline the setup for our simulations.

\subsection{Baryon Acoustic Oscillations}\label{subsec:BAO}

We incorporate the recent BAO measurements reported by the DESI collaboration \cite{DESI:2024mwx}, which span a wide range of cosmic epochs and include approximately six million extragalactic sources. The data set comprises four primary tracers of large-scale structure: the Bright Galaxy Sample (BGS), Luminous Red Galaxies (LRGs), Emission Line Galaxies (ELGs) and Quasars (QSOs). These tracers are distributed across seven observational bins, each corresponding to different cosmic distances, allowing DESI to effectively map the expansion history of the Universe from the relatively nearby Universe to the distant cosmic web. The BGS provides information from the lowest-distance regime, followed by LRGs and ELGs in intermediate intervals. A mixed sample combining high-distance LRGs with low-distance ELGs enhances coverage before the transition region. At even greater distances, additional ELG observations and QSOs extend the reach of the survey. The most distant measurements are obtained from the Lyman-$\alpha$ forest at high redshifts ($z \leq 4.16$), corresponding to a relatively early stage in the evolution of the Universe. This final observation constitutes the highest-redshift measurement among all the observations considered in this work. 

The BAO method relies on the characteristic clustering scale imprinted by acoustic waves in the early Universe, quantified by the sound horizon at the drag epoch, $r_d$. This scale corresponds to the maximum distance travelled by sound waves in the primordial plasma before baryons decoupled from photons and is defined as follows,
\begin{equation}
    r_d = \int_{z_d}^{\infty}\frac{c_s(z)}{H(z)}dz\, , 
\end{equation}
being $c_s(z)$ the sound speed in the baryon-photon fluid and $z_d$ corresponds to the drag epoch redshift computed through the fitting formula given by \cite{Eisenstein:1997ik}. The BAO measurements across the different samples provide observational constraints on $D_V/r_d$, $D_M/r_d$ and $D_H/r_d$. For the BGS and QSO tracers, only the angle-averaged parameter $D_V/r_d$ is extracted due to their limited signal-to-noise ratio. In contrast, the remaining tracers yield separate determinations of $D_M/r_d$ and $D_H/r_d$, together with their correlation coefficient $r$, offering insight into the anisotropic clustering signal. For our analysis, we focus exclusively on $D_V/r_d$ and $D_H/r_d$ measurements, excluding $D_M/r_d$ due to its strong correlation with $D_H/r_d$, which could bias parameter inference. Here, $D_V$ is the isotropic volume-averaged distance defined as:
\begin{equation}
D_V(z) \equiv \left[ z D_H(z) D_M^2(z) \right]^{1/3},
\end{equation}
and $D_H$ is the Hubble distance:
\begin{equation}
D_H(z) \equiv \frac{c}{H(z)},
\end{equation}
where $D_M(z)$ is the comoving distance. For our analysis the chi-square function is computed comparing observed and theoretical distance measurements as:
\begin{equation}
\chi^2_{\rm DESI} = \sum_{i} \left(\frac{D_{\rm obs}(z_i) - D_{\rm model}(z_i)}{\delta D_{\rm obs}(z_i)}\right),
\end{equation}
where $ D_{\rm obs}(z_i) $ represents the observed BAO distances, either $ D_V(z_i)/r_d $ (spherically averaged) for BGS and QSO tracers, or $ D_H(z_i)/r_d $ (Hubble distance) for the remaining tracers, $ D_{\rm model}(z_i) $ are the corresponding theoretical predictions and $\delta D_{\rm obs}(z_i)$ are the reported uncertainties for each redshift bin.


\subsection{Cosmic Microwave Background}\label{subsec:CMB}

The CMB is a relic radiation field that provides a snapshot of the Universe at the epoch of recombination ($z \approx 1100$). Its nearly isotropic blackbody spectrum, with temperature $T_0 = 2.725\,\text{K}$, exhibits minute anisotropies that encode information about the initial conditions of the cosmos. Given our focus on background-level modifications to the standard cosmological model, we adopt the Planck Compressed 2018 dataset \cite{Chen_2019} rather than the full Planck likelihoods. This compressed approach has been used \cite{LinaresCedeno:2020uxx, Shen:2025cjm, Yang:2025boq,Kumar:2025mzo} for constraining both $\Lambda$CDM and its extensions. The cosmological models explored in this work represent theoretically consistent extensions of the standard $\Lambda$CDM paradigm, while preserving the fundamental gravitational framework of General Relativity. This ensures the compressed data remains appropriate for parameter estimation. We constrain the cosmological parameters through the following quantities:
\begin{enumerate}
    \item The \textit{acoustic scale} $\ell_A$, which determines the angular spacing of CMB temperature anisotropies:
    \begin{equation}
        \ell_A = (1+z_*)\frac{\pi D_A(z_*)}{r_s(z_*)}
        \label{eq:acoustic_scale}
    \end{equation}    
    \item The \textit{shift parameter} $\mathcal{R}$, governing the overall amplitude of acoustic peaks:
    \begin{equation}
        \mathcal{R} = (1+z_*) \frac{\Omega^{1/2}_{0m} H_0}{c} D_A(z_*)
        \label{eq:shift_param}
    \end{equation}
\end{enumerate}
Here $z_*$ denotes the redshift at photon decoupling, $r_s(z_*)$ represents the sound horizon at decoupling and $D_A(z_*)$ is the angular diameter distance defined as,
\begin{equation}
D_A(z) = \frac{c}{H_0} \frac{1}{1+z} \int_0^z \frac{dz'}{E(z')}.
\label{eq:DA_flat}
\end{equation}
In addition, the comoving sound horizon is computed through:
\begin{equation}
{r}_s(z) = \int_{0}^{1/(1+z)} \frac{da}{a^2 E(a) \sqrt{1 + \frac{3\Omega_{0b}h^2}{4\Omega_{0\gamma}h^2} a}},
\label{eq:sound_horizon}
\end{equation}
with the photon density parameter given by:
\begin{equation}
\Omega_\gamma h^2 = 2.3809 \times 10^{-5} \left(\frac{2.7\,\text{K}}{T_{\text{CMB}}}\right)^{-4},
\label{eq:omega_gamma}
\end{equation}
with $T_{\text{CMB}}=2.7255 \ \rm{K}$. The statistical comparison between models and Planck data is performed via:
\begin{equation}
\chi^2_{\text{CMB}} = \sum_{i,j} \Delta x_i \cdot (\mathcal{C}^{-1})_{ij} \cdot \Delta x_j,
\label{eq:chi2_distance_priors}
\end{equation}
where $\Delta x_i = \{ \mathcal{R}^{\text{model}} - \mathcal{R}^{\text{Planck}}, \ell_A^{\text{model}} - \ell_A^{\text{Planck}}, \Omega_b h^2_{\text{model}} - \Omega_b h^2_{\text{Planck}} \}$ represents the deviation vector between model predictions and Planck measurements, $\mathcal{C}_{ij}$ denotes the inverse covariance matrix, with two distinct cases. For the $\Lambda$CDM model, we use the covariance matrix derived from $\Lambda$CDM fits to the Planck data. On the other hand, for the remaining models (e.g., $w$CDM, CPL and Chebyshev), we employ the covariance matrix obtained from $w$CDM fits, as these better capture the parameter degeneracies in extensions to the standard paradigm (see \cite{Chen_2019} for details). We note that the CMB shift parameters are not direct observables, but derived quantities obtained from the full CMB power spectrum under the assumption of a fiducial cosmological model, typically $\Lambda$CDM. As a consequence, their use entails a mild degree of model dependence. The cosmological extensions considered in this work are constructed as smooth extensions of $\Lambda$CDM and are not expected to introduce significant modifications to the physics of recombination or to the sound horizon at decoupling.
In particular, deviations from $\Lambda$CDM in the models explored mainly affect the late-time expansion history, while early-universe physics remains effectively unchanged within the parameter space considered. Therefore, the use of CMB shift parameters provides a reliable and computationally efficient approximation to the full CMB likelihood. Under these assumptions, any residual model dependence associated with the shift parameters is expected to be subdominant with respect to the statistical uncertainties of the data.

\subsection{Cosmic Chronometers}\label{subsec:CC}

\begin{equation}
H(z)=-\frac{1}{1+z}\frac{\mathrm{dz}}{\mathrm{dt}},
\end{equation}
where $\mathrm{d}z/\mathrm{d}t$ is estimated through the $4000~\mbox{\AA}$ spectral break, which varies as a function of redshift. This method enables a direct determination of the Hubble parameter by employing spectroscopic dating techniques to compare the ages and metallicities of these galaxies. This expression can be operationalised by estimating $\mathrm{d}z/\mathrm{d}t$ through the ratio $\Delta z / \Delta t$, where $\Delta t$ is the age difference between two passively evolving galaxies that formed simultaneously but are observed at slightly different redshifts, separated by a small interval $\Delta z$. In this work, we adopt CC data from \citep{Magana:2017nfs} comprising a total of 31 data points. In this way, the cosmological parameters can be constrained using the following chi-squared function,
\begin{equation}
\chi_{\rm CC}^2 = \sum_{i=1}^{N_{\rm CC}} \frac{ \left[ H(z_{i}) -H_{\rm obs}(z_{i})\right]^2 }{ \sigma_{H_i}^{2} },
\end{equation}
where $N_{\rm CC}$ is the number of the observational Hubble parameter $H_{\rm obs}(z_{i})$ at $z_{i}$,
$\sigma_{H_i}$ is its error and $H(z_{i})$ represents the theoretical value predicted by a given model at $z_{i}$.

\subsection{Strong Lensing Systems}

Strong gravitational lensing provides an effective method to constrain cosmological parameters through the geometric distortion of background sources. We use the fiducial compilation from \cite{Amante:2019xao} which comprises 143 strong lensing systems (SLS) with elliptical galaxies acting as lenses, covering a redshift range for the lens $0.0625 < z_l < 0.958 $ and for the source $0.2172 < z_s < 3.595$. The singular isothermal sphere (SIS) model predicts the Einstein radius as \cite{2015eaci.book.....S}:
\begin{equation}
\theta_{E} = 4 \pi \frac{\sigma^2_{\text{SIS}}}{c^2} \frac{D_{ls}}{D_s},
\label{thetaESIS}
\end{equation}
    where $\sigma_{\text{SIS}}$ represents the velocity dispersion of the lens, while $D_{ls}$ and $D_s$ correspond to the angular diameter distances (see Eq. \ref{eq:DA_flat}) between lens-source and observer-source respectively and $c$ is the speed of light. By defining the distance ratio  $D \equiv D_{ls}/D_{s}$, it is natural to define the lensing observable and its theoretical counterpart as,
\begin{equation}
D^{\rm obs} = \frac{c^2 \theta_E}{4\pi \sigma^2}, \quad
D^{\rm th}(z_l, z_s; \Theta) = \frac{\int_{z_l}^{z_s} E(z^{\prime},\Theta)^{-1} dz^{\prime}}{\int_0^{z_s} E(z^{\prime},\Theta)^{-1} dz^{\prime}},
\label{Dlens_th}
\end{equation}
under this prescription cosmological parameters $\Theta$ are constrained by minimising:
\begin{equation}
\chi^2_{\rm SL}(\Theta) = \sum_{i=1}^{N_{\rm SL}} \frac{[D^{\rm th}_i(\Theta) - D^{\rm obs}_i]^2}{(\delta D^{\rm obs}_i)^2},
\label{eq:chisquareSL}
\end{equation}
where $N_{\rm SL}$ is the number of lensing systems and $\delta D^{\rm obs}$ accounts for observational uncertainties.

\subsection{Joint constraints}
In this work, we present cosmological constraints derived from the joint analysis of four independent observational probes: BAO, CMB, CC and SLS. The combined constraints are obtained by minimising the total chi-squared function, defined as the sum of the individual chi-squared contributions from each dataset:  
\begin{equation}  
    \chi^2_{\rm total} = \chi^2_{\rm DESI} + \chi^2_{\rm CMB} + \chi^2_{\rm CC} + \chi^2_{\rm SL} ,  
\end{equation}  
where each term accounts for the statistical deviations between theoretical predictions and observational data. This multi-probe approach helps break degeneracies in parameter space and provides tighter cosmological constraints than any single probe alone. It is worth noting that we impose a Gaussian prior on $\Omega_{0m}$ exclusively for the SLS analysis. This prior is constructed using the mean and uncertainty of $\Omega_{0m}$ derived from the combined constraints of BAO, CMB and CC for each cosmological model under consideration. We adopt this approach because $\Omega_{0m}$ cannot be well-constrained by gravitational lensing observations alone due to the strong degeneracy with other parameters (see for example  \cite{2013ApJ...768...39G, 2010AIPC.1241.1118P, Chen_2018, Caminha_2022, Verdugo:2024dgs, 2025CQGra..42d5016A}). 

 \subsection{Simulation setup}

Once the best-fitting cosmological parameters for each model were obtained using the data and methodology discussed in Section~\ref{sec:data}, we used them to create initial conditions and run dark matter-only cosmological simulations\footnote{As it is usual in this type of simulation, the parameter $\Omega_{0r}$ is effectively set to zero, since it scales as $\sim a^{-4}$ and its contribution to the Friedmann equations can be neglected at late times.}.

\subsubsection{Initial Conditions}
\label{sec:initial-conditions}
Since in these dark energy models the Universe is accurately described by standard inflationary physics at very early times~\citep{Planck:2020}, we assume a common primordial power spectrum for all cosmologies. The primordial spectrum is described by a power law of the form
\begin{equation}\label{eq:Pk_primordial}
P_{\rm prim}(k) = A_s \left( \frac{k}{k_*} \right)^{n_s-1},
\end{equation}
where $A_s$ sets the amplitude of the primordial fluctuations, $n_s$ is the spectral index and $k_*$ denotes the pivot scale. These parameters fully characterise the statistical properties of the initial density perturbations generated during inflation and are assumed to be identical for all the cosmological models considered in this work.

In standard cosmological perturbation theory, the primordial power spectrum is related to the linear matter power spectrum through the transfer function $T(k)$ and the linear growth factor $D(z)$, such that
\begin{equation}\label{eq:Pk_lin-Pk_primordial}
P_{\rm lin}(k,z) = P_{\rm prim}(k)\,T^2(k)\,D^2(z).
\end{equation}
The transfer function encodes the effects of early-Universe physics, in particular the radiation--matter transition, and depends primarily on the physical matter density $\Omega_{0m} h^2$ and on the baryon fraction $\Omega_{0b}/\Omega_{0m}$ \citep{Eisenstein:1997ik}. The linear growth factor $D(z)$ describes the subsequent gravitational evolution of perturbations at late times and depends on the background expansion history $H(z)$, including the effects of dark energy.

We adopt the best-fitting inflationary parameters inferred from the Planck 2018 $\Lambda$CDM analysis~\citep{Planck:2020}, using the \texttt{plikHM} likelihood (TT, TE, EE + low-$\ell$ + lowE + lensing), fixing $n_s = 0.966$ and $A_s = 5.1 \times 10^{-9}$. For each cosmological model, the best-fit background parameters are used to compute a model-dependent linear transfer function using the Einstein--Boltzmann solver \textsc{CAMB}\footnote{\url{https://github.com/cmbant/CAMB}}\citep{Lewis:1999bs}. CAMB also provides the $\sigma_8$ parameter at arbitrary redshifts; we list the $z=0$ values in Table~\ref{tab:sigma_params}.

\begin{table}
\centering
\caption{Values of $\sigma_8$ at $z=0$ implied by the linear matter power spectrum of each cosmological model, as computed with \textsc{CAMB} using their best-fitting parameters.}
\label{tab:sigma_params}
\begin{tabular}{lc}
\hline
Model & $\sigma_8(z=0)$ \\
\hline
Planck--$\Lambda$CDM & 0.823 \\
$\Lambda$CDM        & 0.818 \\
$w$CDM              & 0.907 \\
CPL                 & 0.911 \\
Chebyshev           & 0.951 \\
\hline
\end{tabular}
\end{table}

To generate the initial conditions for the $N$-body simulations, we use the \textsc{MUSIC} code\footnote{\url{https://bitbucket.org/ohahn/music/src/}}\citep{Hahn:2011}, which generates the linear density field in Fourier space using as input the primordial spectral index, the transfer function and the corresponding normalisation. In particular, the initial density field is generated according to
\begin{equation}
\delta^{(1)}(\mathbf{k},z) = \sqrt{P_{\rm lin}(k,z)}\,\mu(\mathbf{k}),
\end{equation}
where $P_{\rm lin}$ is given by Eq.~\eqref{eq:Pk_lin-Pk_primordial}, $\mu(\mathbf{k})$ is a Gaussian random field with zero mean and unit variance. To generate the initial conditions, in this work we set $z \equiv z_{\rm ini} = 50$, adopt the $\sigma_8(z=0)$ values listed in Table~\ref{tab:sigma_params} and use the best-fit background parameters reported in Table~\ref{tab:parametros_cosmologicos}, which are later used to construct the initial particle positions and velocities via Lagrangian Perturbation Theory (LPT) for each cosmology.

MUSIC allows one to choose among $\Lambda$CDM, $w$CDM and CPL cosmologies by specifying the $w_0$ and $w_1$ parameters of the CPL parametrisation. For the Chebyshev model, we fix the best-fitting coefficients $C_i$ given in Table~\ref{tab:parametros_cosmologicos} and generate the initial conditions assuming a $\Lambda$CDM cosmology (i.e., $w_0=-1$ and $w_1=0$). This choice is motivated by the fact that, at high redshift, the exponential factor in Eq.~(\ref{eq:w2}), becomes vanishingly small even when evaluated using the best-fitting coefficients $C_i$, causing the EoS to approach $-1$, i.e. $\tilde{w}_2(z=50)\approx-1$. Consequently, within the CPL parametrisation implemented in MUSIC, we effectively have $w_0\approx -1$ and $w_1\approx 0$. This justifies the use of the $\Lambda$CDM EoS for generating the initial conditions at $z=50$.


\subsubsection{N-body simulations}
To run our cosmological models, we use and modify the well-established and massively parallelised $N$-body and hydrodynamical code \textsc{RAMSES}\footnote{\url{https://github.com/ramses-organisation/ramses}} \citep{Teyssier:2002}, with the initial conditions described above.

In particular, we modify the routine that integrates the \textit{first} Friedmann equation (Eq.~\ref{eq:lcdm}). This routine defines two functions representing the Friedmann equation, with the Hubble parameter expressed in terms of both the cosmological time $t$ and the supercomoving (or ``conformal''\footnote{\textsc{RAMSES} employs supercomoving coordinates; here, ``conformal time'' refers to the superconformal time, see \citet{Teyssier:2002}.}) time $\tilde{\tau}$. \textsc{RAMSES} then numerically integrates the Friedmann equation to obtain the scale factor $a(\tilde{\tau})$, generating a lookup table that is later used to interpolate quantities at arbitrary times when computing particle positions and velocities.

We added a module implementing the modified Friedmann equations corresponding to our four models of interest, which is called by \textsc{RAMSES} to generate this table. Finally, we implemented a simple parameter control scheme to select the cosmological model and define the parameter values used in the simulations.

For the $N$-body simulations, we use a box size of $100~\mathrm{Mpc}/h$ containing $256^3$ particles, allowing access to modes ranging from the fundamental to the Nyquist frequency, $k \in [0.062,\, 8.04]~h/\mathrm{Mpc}$. \textsc{RAMSES} uses the Adaptive Mesh Refinement (AMR) method, which refines the simulation grid by increasing the number of cells in denser regions. For the smallest scales resolved in \textsc{RAMSES}, we set \texttt{levelmax=18}, corresponding to $2^{18}$ cells along each dimension, thus resolving scales as small as $\approx 3.8$~kpc$/h$. To ensure a consistent comparison, we adopt the same specifications and random seed for all models studied.

In addition to the four models constrained in this paper, we ran a reference cosmological simulation using the best-fitting parameters from the baseline Planck 2018 $\Lambda$CDM analysis, hereafter referred to as Planck--$\Lambda$CDM. This model is based on the same \texttt{plikHM} likelihood mentioned above and is used as a baseline for comparison. The cosmological parameters are $h= 0.673$, $\Omega_{0\rm CDM} = 0.265$, $\Omega_{0\rm b} = 0.049$ and $\Omega_{0\Lambda} = 0.686$.

To summarise, we ran five cosmological simulations using the same numerical setup and pipeline but with parameters appropriate to each model, referred to in this article as Planck--$\Lambda$CDM, $\Lambda$CDM, $w$CDM, CPL and Chebyshev. We emphasise that the last four correspond to cosmologies that adopt the best-fitting parameters from the constraints presented in Section~\ref{sec:Res}, obtained using the observational dataset and methodology described in Section~\ref{sec:data}.

Finally, we note that this simulation setup is limited by its box size, number of particles and the use of a single realisation, which do not allow us to draw definitive conclusions regarding the impact of these models on structure formation or their observational viability in terms of large-scale structure probes. Rather, this setup is intended as a first step towards identifying qualitative and quantitative differences in structure formation among the models in a controlled numerical environment.\\


\section{Results} \label{sec:Res}

\begin{table*}
    \centering
    \renewcommand{\arraystretch}{1.6} 
    \begin{tabular}{l c c c c}
        \toprule
        Parameters / Model & $\Lambda$CDM & $w$CDM & CPL & Chebyshev \\
        \midrule
        $\Omega_{0\textrm{CDM}}$
            & $0.251^{+0.011}_{-0.010}$
            & $0.241^{+0.011}_{-0.011}$
            & $0.264^{+0.017}_{-0.017}$
            & $0.254^{+0.018}_{-0.018}$ \\
        $\Omega_{0b}$
            & $0.053^{+0.002}_{-0.002}$
            & $0.045^{+0.004}_{-0.004}$
            & $0.047^{+0.004}_{-0.004}$
            & $0.042^{+0.005}_{-0.005}$ \\
        $\Omega_{0\textrm{DE}}$
            & $0.696^{+0.011}_{-0.011}$
            & $0.713^{+0.013}_{-0.012}$
            & $0.689^{+0.017}_{-0.017}$
            & $0.704^{+0.021}_{-0.021}$ \\
        $\Omega_{0r}$
            & $8.54^{+0.32}_{-0.32} \cdot 10^{-5}$
            & $7.83^{+0.42}_{-0.42} \cdot 10^{-5}$
            & $8.43^{+0.56}_{-0.55} \cdot 10^{-5}$
            & $8.07^{+0.61}_{-0.62} \cdot 10^{-5}$ \\
        $h$
            & $0.699^{+0.013}_{-0.013}$
            & $0.730^{+0.020}_{-0.019}$
            & $0.703^{+0.024}_{-0.022}$
            & $0.719^{+0.029}_{-0.026}$ \\
        \midrule
        $w_0$
            & $-1$
            & $-1.224^{+0.092}_{-0.100}$
            & $-0.863^{+0.258}_{-0.256}$
            & $-1.232^{+0.328}_{-0.333}$ \\
        $w_1$
            & \textemdash
            & \textemdash
            & $-1.474^{+1.025}_{-1.100}$
            & \textemdash \\
        \midrule
        $C_0$
            & \textemdash & \textemdash & \textemdash
            & $1.809^{+0.673}_{-0.723}$ \\
        $C_1$
            & \textemdash & \textemdash & \textemdash
            & $0.350^{+0.902}_{-0.904}$ \\
        $C_2$
            & \textemdash & \textemdash & \textemdash
            & $-0.530^{+0.684}_{-0.679}$ \\
        $C_3$
            & \textemdash & \textemdash & \textemdash
            & $-0.303^{+0.415}_{-0.480}$ \\
        \bottomrule
    \end{tabular}
    \caption{Constraints on cosmological parameters: posterior means and $1\sigma$ confidence intervals for all models. For the Chebyshev model, the $w_0$ parameter is computed from Eq.~(\ref{eq:w1}), where $\tilde{w}_1(0) \equiv w_0$.}
    \label{tab:parametros_cosmologicos}
\end{table*}

The results of this study are presented in two main parts. The first part presented in subsection \ref{subsec:backgound_const}, focuses on the outcomes of a Bayesian statistical analysis aimed at constraining the previous cosmological models using a joint analysis of four different observational data sets. Specifically, we use observations from BAO, CMB, CC and SLS. This part provides a comparative assessment of the parameter estimations and their uncertainties, allowing us to evaluate the consistency and viability of each model in light of current data. The second part presented in subsection  \ref{subsec:non-linear}, builds upon these constraints to investigate the implications of each model for cosmic structure formation. Specifically, we analyse how the inferred background dynamics influence the growth of matter perturbations, providing further insight into the physical viability and predictive power of the models.

\subsection{Observational Constraints on Background Cosmologies}\label{subsec:backgound_const}

We present constraints on the cosmological parameters for each model through a joint Bayesian analysis assuming a Gaussian likelihood $\mathcal{L}(\theta) = \exp\left(-\frac{1}{2}\chi^2_{\rm total}(\theta)\right)$ incorporating four independent observational probes BAO, CMB, CC and SLS. Our analysis employs a Markov Chain Monte Carlo (MCMC) approach implemented with the \texttt{emcee} package~\cite{Foreman-Mackey:2012any}, using 1000 walkers and running chains for 5000 iterations after convergence (assessed through Gelman-Rubin statistics $R < 1.01$). 

All cosmological models in our analysis share the fundamental $\Lambda$CDM parameters: the baryon density $\Omega_{0b}$, cold dark matter density $\Omega_{0\textrm{CDM}}$ and dimensionless Hubble parameter $h$. We adopt conservative flat priors for these baseline parameters: $\Omega_{0b} \in [0.01, 0.1]$, $\Omega_{\textrm{CDM}} \in [0.15, 0.35]$ and $h \in [0.5, 0.8]$, encompassing current observational constraints. For extended models, we introduce additional parameters with physically motivated priors:

\begin{itemize}
    \item $w$CDM model : The constant dark energy equation of state follows $w_0 \in [-3, 1]$ to encompass both quintessence and phantom energy regimes.
    
    \item CPL: The dark energy EoS parameters $w_0$ and $w_1$ follow wide flat priors $w_0 \in [-3, 1]$, $w_1 \in [-5, 5]$ being the same prior for $w_0$ as in the $w$CDM case.
    
    \item Chebyshev expansion: The coefficients $C_{0-3}$ are assigned Gaussian priors $\mathcal{N}(\mu_{\Lambda \textrm{CDM}}, \sigma=3)$, where $\mu_{\Lambda \textrm{CDM}} = (1, 0, 0, 0)$ for $(C_0, C_1, C_2, C_3)$ respectively. This centres the expansion on the $\Lambda$CDM prediction while allowing data-driven deviations, following the methodology of \cite{DESI:2024aqx}.
\end{itemize} 

\begin{figure*}
    \centering
    
    \includegraphics[width=.8\textwidth]{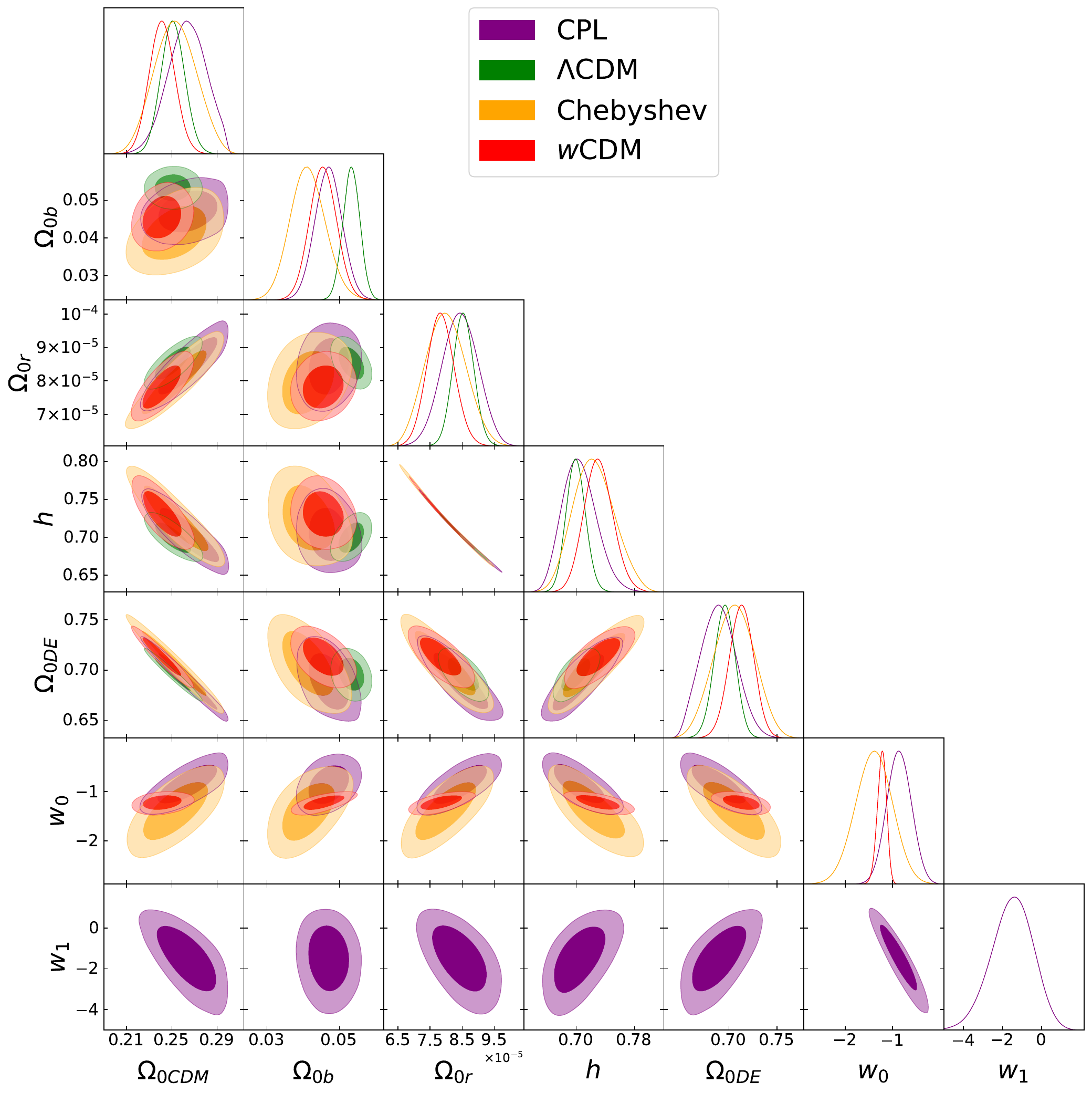}
    \caption{1D marginalised posterior distributions and the 2D 68\% and 95\%  confidence levels for the parameters of the four models considered in this work.}
    \label{jointCont}
\end{figure*}

\begin{figure}
    \centering
    \includegraphics[width=0.5\textwidth]{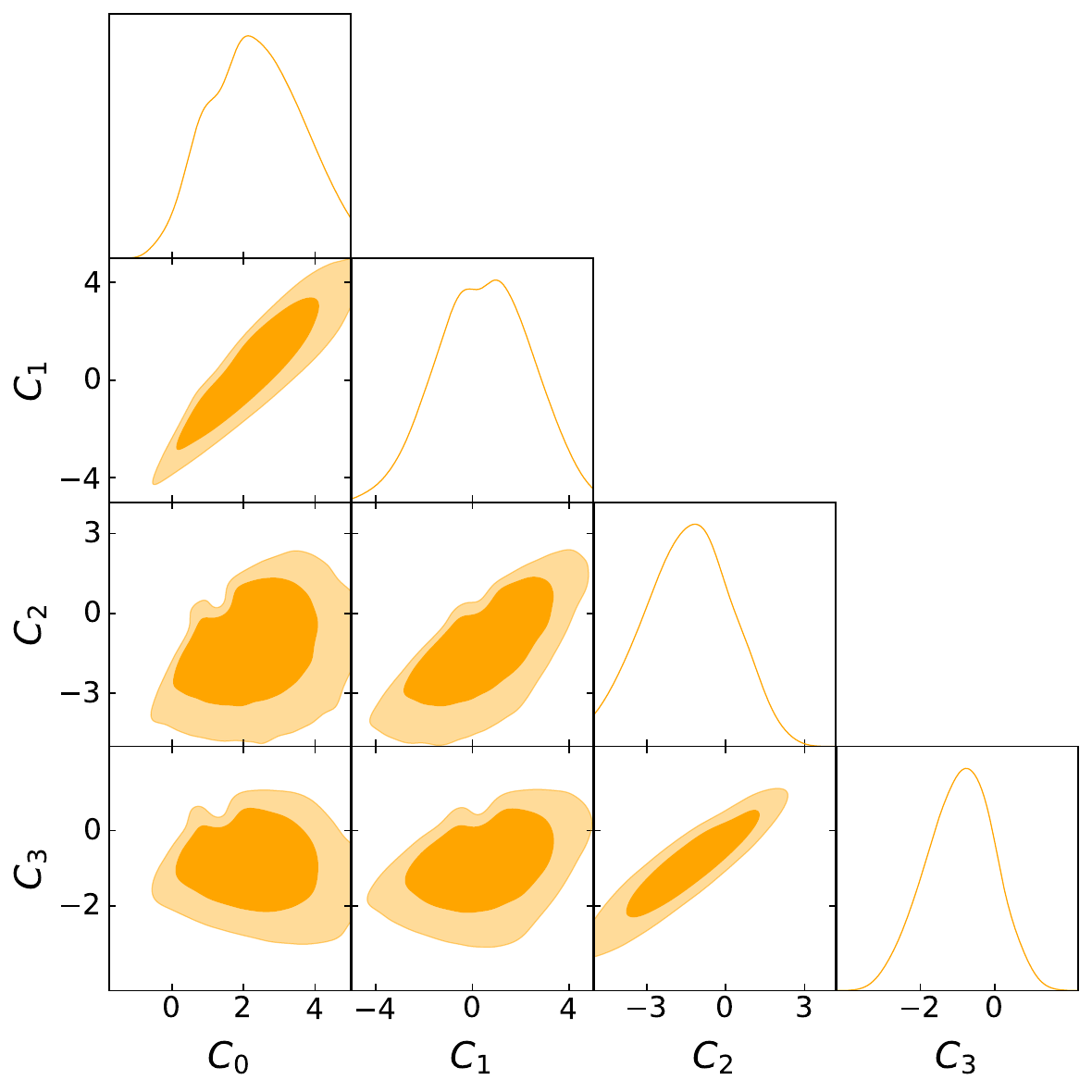}
    \caption{1D marginalised posterior distributions and the 2D 68\% and 95\%  confidence levels for the Chebyshev coefficients.}
    \label{chebcof}
\end{figure}

Fig.~\ref{jointCont} display the statistical confidence contours for the cosmological parameters of each model under consideration. Aditionally, Fig.~\ref{chebcof} shows the confidence contours for the Chebyshev coefficients $C_0$, $C_1$, $C_2$ and $C_3$. The corresponding posterior mean values and their associated 1$\sigma$ uncertainties (constrained by the joint analysis) are summarised in Table~\ref{tab:parametros_cosmologicos}. Overall, the baseline parameters $\Omega_{0\textrm{CDM}}$, $\Omega_{0b}$ and $\Omega_{0\textrm{DE}}$ remain broadly consistent across all models. The dimensionless Hubble parameter $h$ exhibits slightly more dispersion, with the $w$CDM model favouring a marginally higher value compared to $\Lambda$CDM.

\begin{figure*}
    \centering
    \includegraphics[width=1.0\textwidth]{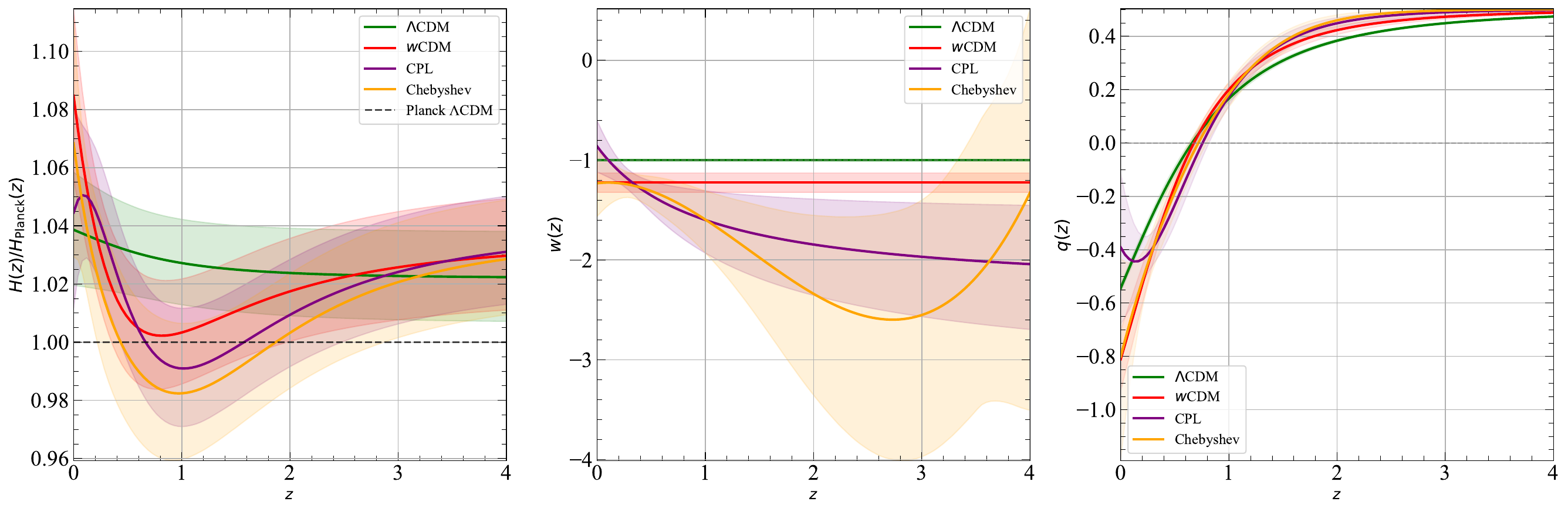} 
    \caption{
A panel of three plots showing the redshift evolution of the Hubble parameter normalised by the Planck reference, $H(z)/H_{\rm Planck}(z)$ (left column), the dark energy equation of state $w(z)$ (middle column) and the deceleration parameter $ 
q(z) = (1+z) \frac{1}{E(z)} \frac{dE(z)}{dz} - 1
$ (right column), for the cosmological models under consideration. The shaded regions represent the $1\sigma$ confidence intervals derived from error propagation of the chains. The results are based on the parameter constraints obtained from a joint fit to all observational data sets considered in this work.}
\label{fig:plots_background}
\end{figure*}

Fig.~\ref{fig:plots_background} shows the reconstruction of the Hubble parameter normalised by Planck reference, the deceleration parameter $q(z)$ and the dark energy EoS parameter.  The dark energy EoS parameter at present time show more distinctive behaviour. For the $w$CDM model, we find $w_0 = -1.224^{+0.092}_{-0.100}$, favouring a phantom-like dark energy component. The CPL parametrization yields $w_0 = -0.8628^{+0.2582}_{-0.2560}$ and $w_1 = -1.4743^{+1.0246}_{-1.1}$, consistent with a dynamical evolution but with larger uncertainties. For the Chebyshev model, the derived present-day value is $w_0 = -1.232^{+0.328}_{-0.333}$, while the coefficients $C_0$ to $C_3$ show no strong evidence ($>2\sigma$) for deviations from their $\Lambda$CDM-centered priors. The behavior of the deceleration parameter $q(z)$ shows that, in the current epoch, all models considered undergo a phase of accelerated expansion.

The transition redshift $z_t$ was determined for all cosmological models. 
For the $\Lambda$CDM model, we find 
$z_t = 0.659^{+0.028}_{-0.028}$.  The $w$CDM model yields a consistent value of  $z_t = 0.671^{+0.027}_{-0.026}$, fully consistent with $\Lambda$CDM within the quoted uncertainties.  In contrast, the CPL parameterisation gives  $z_t = 0.745^{+0.053}_{-0.063}$,  indicating an earlier transition epoch at the $\sim 1.3\sigma$ level. Finally, the Chebyshev model presents $z_t = 0.704^{+0.078}_{-0.086}$.  These results suggest that while dynamical dark energy models allow for moderate variations in the transition epoch, they remain statistically consistent with the $\Lambda$CDM prediction (see Fig.~\ref{fig:plots_background}).

Overall, while the $\Lambda$CDM model remains consistent with constraints from Planck-$\Lambda$CDM, the extended models introduce additional degrees of freedom that allow for modest deviations in key parameters such as $\Omega_{0\textrm{DE}}$, $h$, $\Omega_{0b}$ and 
$w_0$. These deviations may provide valuable insights into the nature of dark energy and potential resolutions to current cosmological tensions. 
Furthermore, the subsequent sections explore the potential effects of these extended model parameters on structure formation, aiming to gain insight into their possible impact.\\

\subsection{Structure formation}\label{subsec:non-linear}
We now explore the structure formation in each 
cosmological model. For this, we run a set of N-body simulations using the best-fitting values for the parameters of each model, as obtained from the observational background constraints in Section~\ref{subsec:backgound_const}.

We start by showing the projected particle mass of the full box at redshift $z=0$, incorporating a zoom-in region of 15 Mpc/$h$ (Fig.~\ref{fig:density_zoom}), then the matter power spectrum (Figure \ref{fig:power_spectra}), the halo mass function (HMF) (Fig.~\ref{fig:halo_mass_function}), the density profile for the most massive haloes of each simulation (Fig.~\ref{fig:shell_profiles_most_massive}) and the averaged density profiles for mass intervals at different redshifts in Figs.\ref{fig:dens_profiles_mean_highz} and \ref{fig:dens_profiles_mean_lowz}. 

\subsubsection{Visual inspection}

\begin{figure*}
    \centering
    \includegraphics[width=\linewidth]{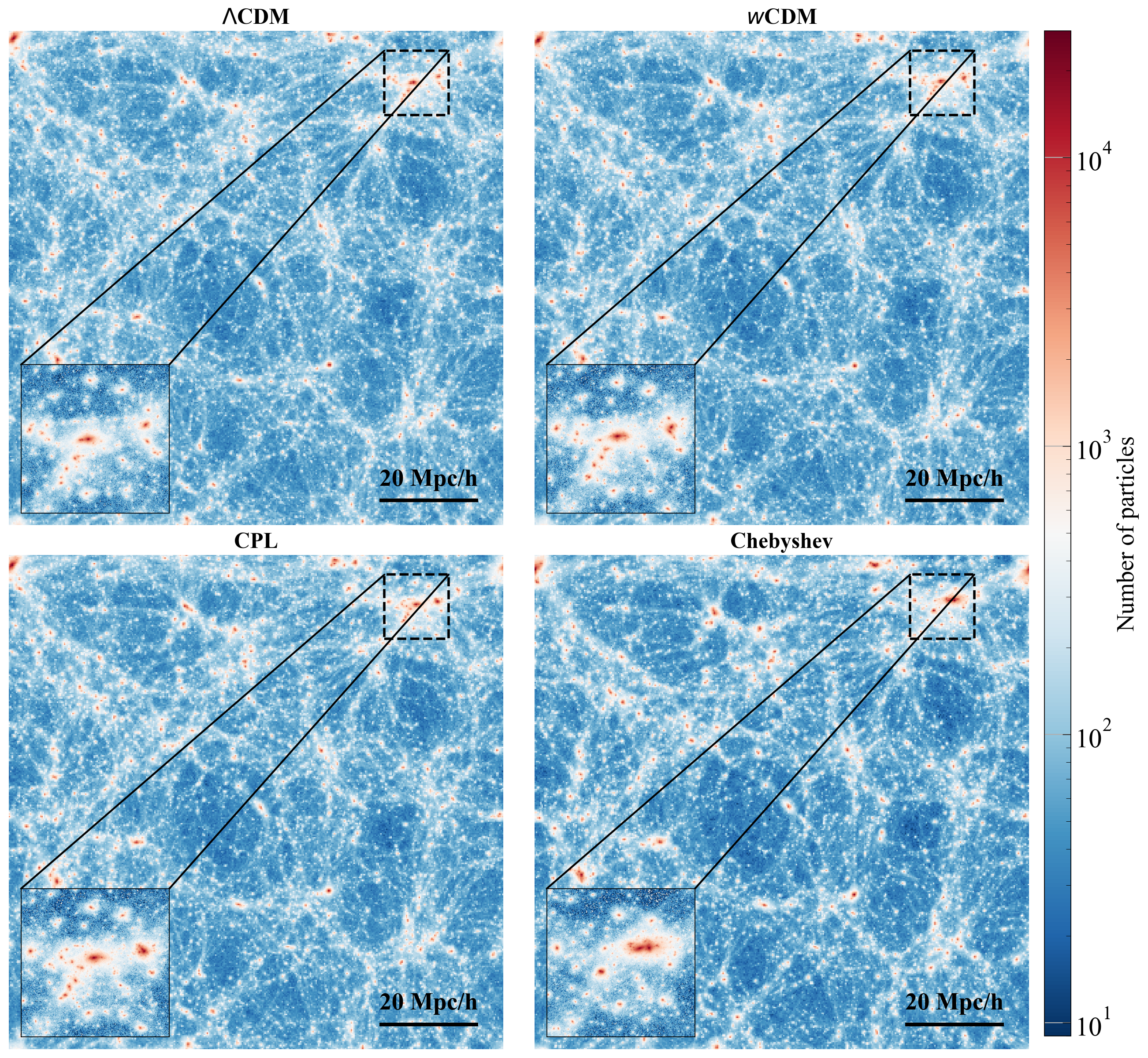}
    \caption{Projected density map along the $z$-axis for a $100~\mathrm{Mpc}/h$ region simulated under the $\Lambda$CDM cosmology at $z=0$. A $20~\mathrm{Mpc}/h$ scale bar is shown in the bottom-right corner. The red square highlights the zoomed-in region, which is displayed in the adjacent panel. The colourbar represents particle number.}
    \label{fig:density_zoom}
\end{figure*}

Fig.~\ref{fig:density_zoom}, shows a two-dimensional projection 
of the dark matter particles along the $z$-axis over the full box ($100$ Mpc/$h$), taken from snapshots of the $\Lambda$CDM,  $w$CDM, CPL and Chebyshev simulations. The black square marks a region of $15$~Mpc/$h$, which is zoomed in  to reveal finer details of the local clustering of matter and halo substructure. The colour bar indicates the number of particles, highlighting the most massive haloes in the simulations. Also, the difference in colour contrast highlights the distinction between filaments and voids, with the Chebyshev simulation showing the most pronounced structures.

In the zoomed-in region, we observe qualitative differences among the models. 
The $\Lambda$CDM simulation displays two distinctive sub-regions characterised by the most massive haloes, surrounded by less massive structures. 
The $w$CDM model shows a broadly similar large-scale distribution, with slightly enhanced local clustering at these scales.
The CPL model, on the other hand, shows even stronger clustering in the same region, resulting in the formation of two more massive haloes. Finally, the Chebyshev model illustrates a scenario in which these two haloes are nearly merging, resulting in an even more massive structure. Furthermore, the Chebyshev model appears to reduce the overall number of haloes in the zoomed-in region, 
since structures in this model are further along in the merging process. This effect can also be observed in the full $100$ Mpc/$h$ simulation box, where increased merging activity is observed in certain regions.

We stress that this comparison is based on a single region at a fixed redshift and is therefore intended as an illustrative visual inspection rather than a statistical statement. 
A quantitative assessment of clustering differences is presented in the following subsections.

\subsubsection{Power spectra}


\begin{figure*}
    \centering
    \includegraphics[width=.90\linewidth]{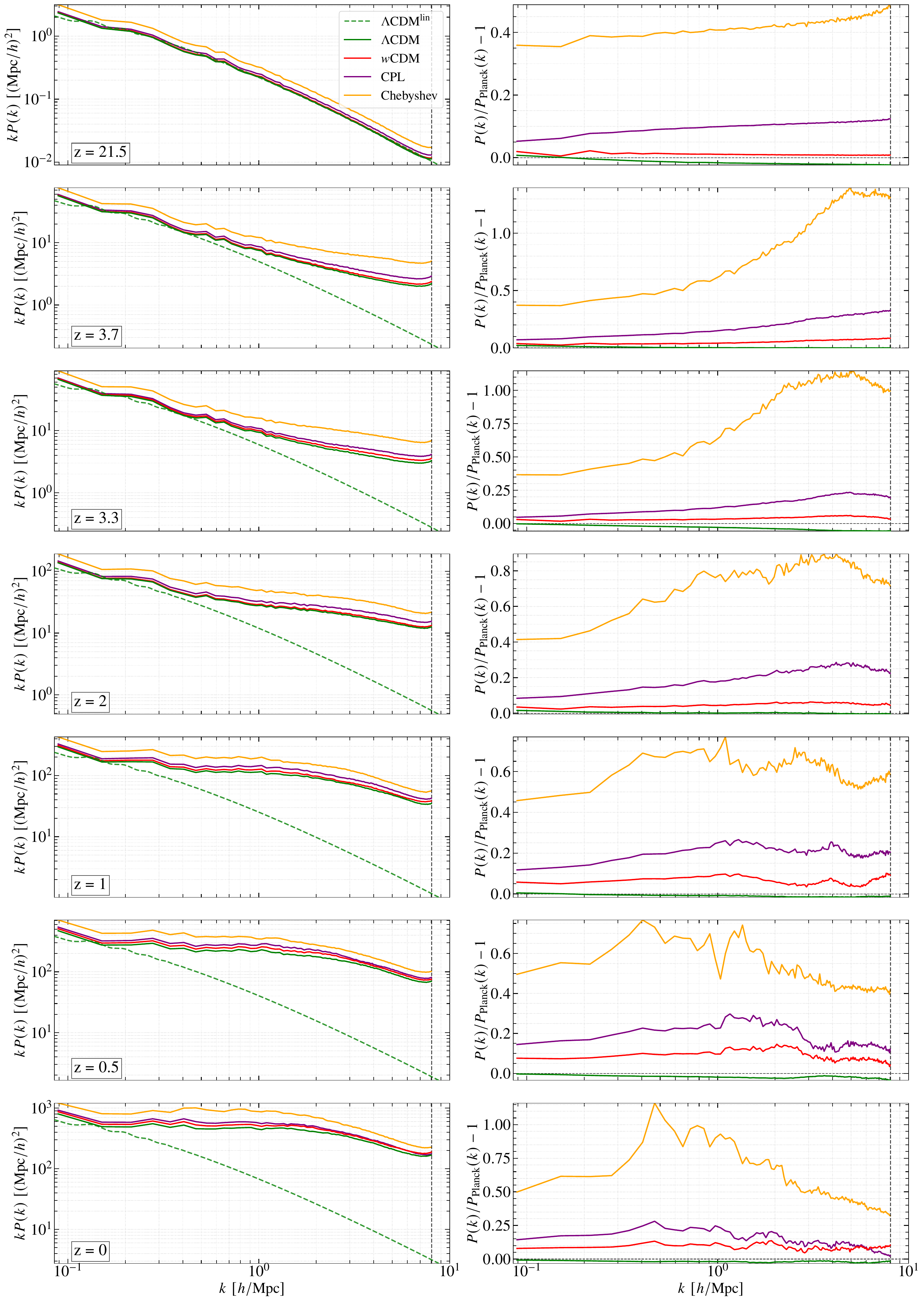}
    \caption{Left column: matter power spectrum 
    measured from the simulations, shown as the combination $kP(k)$, which visually enhances the difference between models. 
    The green, red, purple and orange colours correspond to the $\Lambda$CDM, $w$CDM, CPL and Chebyshev cosmologies, respectively. Each power spectrum is shown up to the Nyquist frequency, $k_N=8.04$~$h$/Mpc. 
    Right column: relative difference between the power spectra and the Planck-$\Lambda$CDM reference spectrum, ${P(k)}/{P(k)}_{\mathrm{\rm Planck}} - 1$. The various rows show the results at redshifts $z = 21.5,\ 3.7,\ 3.3,\ 2.0,\ 1,\ 0.5,$ and $0$, from top to bottom.}
    \label{fig:power_spectra}
\end{figure*}

Fig.~\ref{fig:power_spectra} shows the matter power spectra 
measured from the simulations, shown as solid lines. The green, red, purple and orange colours correspond to the $\Lambda$CDM, $w$CDM, CPL and Chebyshev cosmologies, respectively. In addition, we show the linear matter power spectrum for $\Lambda$CDM as a reference, shown as a green dashed line. Each row shows results for the redshifts $z = 21.5,\ 3.7,\ 3.3,\ 2.0,\ 1,\ 0.5,$ and $0$, from top to bottom. 
Each power spectrum is shown up to the Nyquist frequency, $k_N=8.04$~$h$/Mpc. 
The left column displays $kP(k)$, which visually enhances the difference between models and the right column shows the relative difference 
relative to the Planck-$\Lambda$CDM reference spectrum, ${P(k)}/{P(k)}_{\mathrm{\rm Planck}} - 1$.


Linear perturbation theory predicts a scale-independent growth of matter perturbations in the $\Lambda$CDM model, i.e., all modes grow proportionally to a single growth factor $D(z)$. This is reflected in the power spectrum at high redshift, where evolution remains linear, as seen at $z = 21.5$. Here, the $\Lambda$CDM and $w$CDM models remain consistent with the $\Lambda$CDM linear prediction, while the CPL and Chebyshev models exhibit larger power spectrum amplitudes, with the Chebyshev model showing the largest deviation, therefore, indicating faster structure formation at the linear level. 

At $z = 3.7$, we find that the linear approximation has broken down and scale dependence becomes apparent: small-scale modes begin to grow faster due to mode coupling, while large-scale modes grow more slowly or stabilise in the extended models with respect to the Planck-$\Lambda$CDM reference. This behaviour marks an earlier and clearer transition to the non-linear regime in the CPL and Chebyshev models.

At redshift $z=3.3$, the power enhancement relative to the Planck-$\Lambda$CDM reference begins to extend towards larger scales (i.e. smaller $k$) for the extended models. In particular, the CPL and Chebyshev models exhibit an enhanced growth rate at intermediate scales ($0.3 \lesssim k \lesssim 3$~$h$/Mpc), while there is a slight improvement at the same scales in the $w$CDM case.
This behaviour becomes even more pronounced at $z \leq 2$, indicating a scale-dependent growth in which large-scale modes exhibit a higher growth rate than small-scale modes relative to the Planck--$\Lambda$CDM reference. In contrast, the $\Lambda$CDM model (green solid line) remains consistent with the Planck reference over the full range of scales and redshifts. 

As discussed in Section~\ref{sec:initial-conditions}, although all cosmologies use the same parametrisation of the primordial power spectrum, Eq.~\eqref{eq:Pk_primordial}, and share identical values of $(A_s,n_s)$, their linear matter power spectra---related to the former via Eq.~\eqref{eq:Pk_lin-Pk_primordial}---differ because the physical matter density $\Omega_{0m} h^2$ varies across models (see Table~\ref{tab:parametros_cosmologicos}). Since matter--radiation equality occurs at $1+z_{\rm eq} = (\Omega_{0m} h^2)/(\Omega_{0r} h^2)$ and $\Omega_{0r} h^2$ is fixed by the CMB temperature, variations in $\Omega_{0m} h^2$ shift matter--radiation equality to higher redshift when $\Omega_{0m} h^2$ is larger. This increases the equality horizon scale, $k_{\rm eq} \propto \Omega_{0m} h^2$ \citep{Eisenstein:1997ik}, thereby reducing the range of modes that enter the horizon during radiation domination and enhancing the linear power spectrum amplitude already at high redshift.

In particular, the $w$CDM model exhibits a larger $\Omega_{0m} h^2$ than the Planck--$\Lambda$CDM case (Table~\ref{tab:parametros_cosmologicos}), which enhances the linear amplitude through the transfer function and contributes to its larger $\sigma_8$ value (Table~\ref{tab:sigma_params}). Despite its higher late-time expansion rate relative to Planck--$\Lambda$CDM (Fig.~\ref{fig:plots_background}), this early-time enhancement dominates the final amplitude. The CPL and Chebyshev models also present slightly larger values of $\Omega_{0m} h^2$, which induce a comparable but subdominant early-time enhancement. However, in these cases additional differences arise from their modified background expansion histories $H(z)$, which alter the linear growth factor $D(z)$ and amplify the late-time evolution of density perturbations. This cumulative growth effect is particularly relevant for the Chebyshev model, where it drives the largest deviation in $\sigma_8$ relative to the Planck--$\Lambda$CDM reference.

\subsubsection{Halo abundances}

We now present the results of the halo mass function (HMF), defined by ${\rm d}n/{\rm d}\ln M$, where $n$ is the number density of haloes and $M\equiv M_{\rm 200c}$ is the halo mass in units of $M_\odot/h$. To build the halo catalogues from the simulations, we used the AMIGA Halo Finder (AHF) code~\citep{knollmann:2009}\footnote{\url{http://popia.ft.uam.es/AHF/}}. AHF identifies haloes by applying a spherical overdensity criterion based on the mean enclosed density relative to the critical density $\rho_c(z)$. As the latter depends on the cosmology, we supplied AHF with tables of $H(z)$ and $\Omega_m(z)$ for each model via the \texttt{DarkEnergyFile} parameter. In all runs, we fixed

$$
\Delta = 200,~~N_{\min} = 20~\text{particles per halo},~~\texttt{VescTune} = 1.5,
$$
(i.e.\ 1.5 × the nominal escape velocity). For the halo‐finding grid we set $\texttt{LgridDomain} = 256$ and $\texttt{LgridMax} = 256^3$).

\begin{figure*}
    \centering
    \includegraphics[width=.9\linewidth]{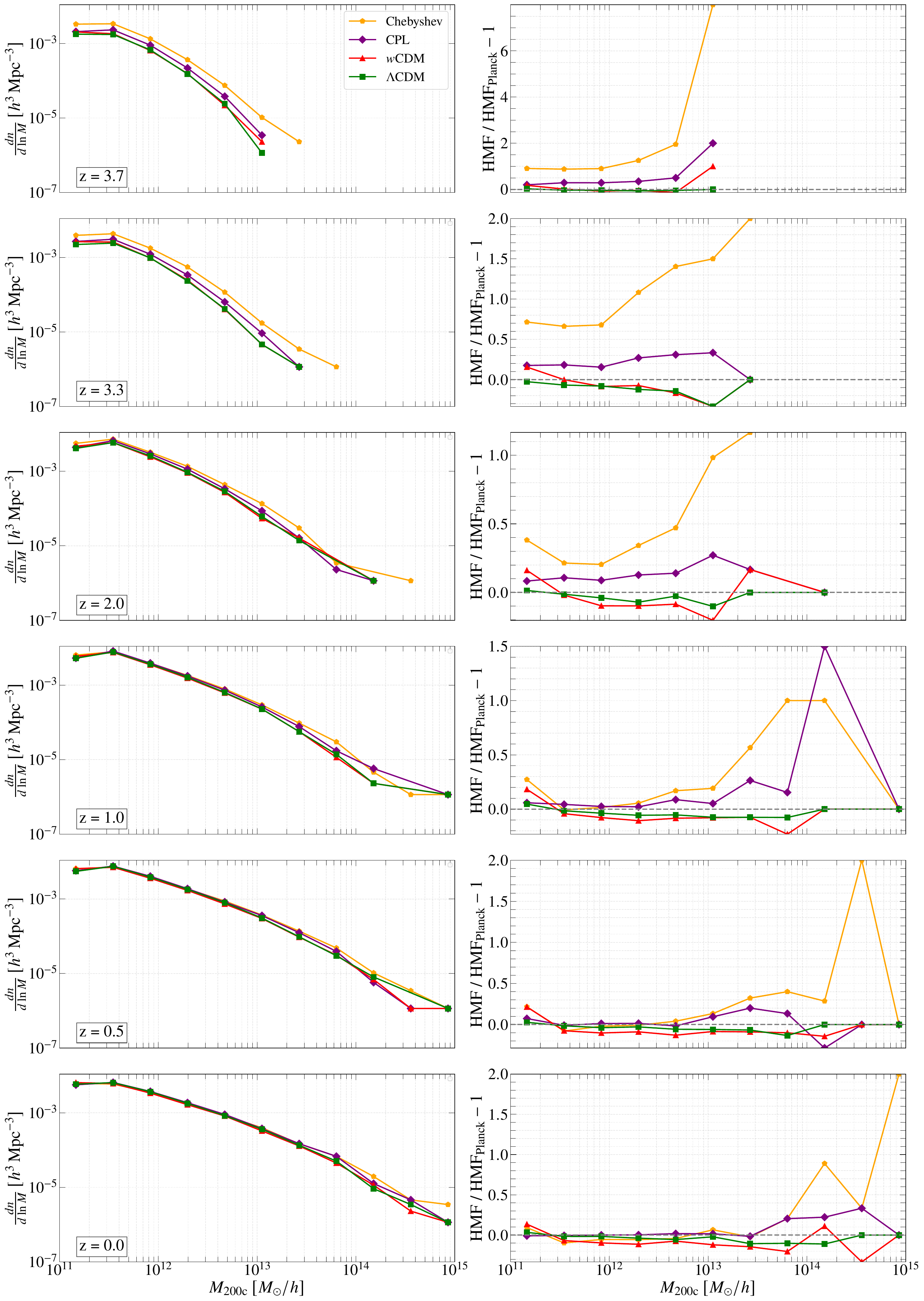}
    \caption{Left column: halo mass function (HMF) measured from the simulations. 
The green, red, purple and orange colours correspond to the $\Lambda$CDM, $w$CDM, CPL and Chebyshev cosmologies, respectively. Right column: relative difference between the HMF of each model and that of Planck-$\Lambda$CDM; $\mathrm{HMF}/\mathrm{HMF}_{\mathrm{\rm Planck}} - 1$. The various rows show results for the redshifts
$z = 3.7,\ 3.3,\ 2.0,\ 1,\ 0.5,$ and $0$, from top to bottom.}
    \label{fig:halo_mass_function}
\end{figure*}

Given the simulation box size of $100$ Mpc/$h$, we do not expect to obtain robust statistics for the most massive haloes, as the limited volume suppresses the large-scale modes required to form galaxy clusters. In addition, in some models, no haloes are found in specific mass bins at a given redshift; for clarity, we exclude these bins from the visual presentation when we show the relative difference regarding the Planck-$\Lambda$CDM simulation in the right column of Fig.~\ref{fig:halo_mass_function}.

Fig.~\ref{fig:halo_mass_function} shows the HMF for all models, where green, red, purple and orange colours correspond to the $\Lambda$CDM, $w$CDM, CPL and Chebyshev cosmologies, respectively. The right column shows the relative difference between the HMF of each model and that of Planck-$\Lambda$CDM; $\mathrm{HMF}/\mathrm{HMF}_{\mathrm{\rm Planck}} - 1$. Each row shows the redshifts
$z = 3.7,\ 3.3,\ 2.0,\ 1,\ 0.5 $ and $0$.

At high redshift ($z\geq2 $), both the CPL and Chebyshev models exhibit an enhancement in the HMF compared to the Planck-$\Lambda$CDM case at the lower mass range, ($11 \lesssim \log M_{\rm 200c} \lesssim 13$) 
indicating an earlier formation of structures in these cosmologies. The Chebyshev model, in particular, not only increases the overall abundance of haloes but also leads to the formation of significantly more massive haloes at early times, as is shown in the left column.

In contrast, the $w$CDM model exhibits a mild enhancement at the low-mass end but a suppression towards higher masses at $z\leq3.3$. Although this cosmology displays an enhanced power spectrum amplitude relative to the Planck--$\Lambda$CDM reference, halo abundance depends not only on the fluctuation amplitude but also on the subsequent growth history and on the mass definition adopted. The smaller matter density parameter $\Omega_{0m}$ (see Table~\ref{tab:parametros_cosmologicos}) and the higher expansion rate $H(z)$ (Fig.~\ref{fig:plots_background}) modify the late-time growth of matter perturbations through the evolution of $\Omega_m(z)\equiv\rho_m(z)/\rho_{\rm c}(z)$ and through the Hubble friction term $2 H \dot{\delta}$ in the linear growth equation $\ddot{\delta} + 2 H \dot{\delta} - 4 \pi G \rho_m \delta = 0$, thereby influencing the formation of the most massive systems. Moreover, since halo masses are defined as $M_{\rm 200c}$, the larger critical density $\rho_c(z)\propto H^2(z)$ leads to smaller characteristic radii $R_{\rm 200c}$ and hence smaller enclosed masses for a given physical halo. The combined effect of these factors can therefore reduce the abundance of haloes above fixed mass thresholds, particularly at the high-mass end.

As expected, the $\Lambda$CDM model shows good agreement with the Planck-$\Lambda$CDM reference, with only a mild suppression developing at $z \leq 3.3$, reaching $\sim 10\%$ at $z=0$.

At low redshift $(z\leq1)$, the CPL and Chebyshev models again exhibit an enhancement in the HMF compared to the Planck-$\Lambda$CDM case. However, this enhancement now appears in the higher mass range  ($13 \lesssim \log M_{\rm 200c} \lesssim 15$),  while the lower mass range shows better agreement with Planck-$\Lambda$CDM. That is, at low redshift, the formation rates of lower mass haloes start to decrease, whereas the formation rate of more massive haloes increases. 

It is worth noting that even with these differences in halo formation rate, the Chebyshev model, at $z=0$, still shows a slight increase in the number of lowest-mass haloes, while it is roughly consistent with $\Lambda$CDM at intermediate masses ($12 \lesssim \log M_{\rm 200c} \lesssim 13$) within 10$\%$ (where we see a slight suppression). At this redshift, the Chebyshev model also shows a significantly higher number of massive haloes---up to twice more than Planck-$\Lambda$CDM in the most massive bins analysed. This overabundance can be due to the merger of smaller haloes, as suggested by the aforementioned suppression in the lower mass range with respect to Planck-$\Lambda$CDM. Overall, among all the models considered, Chebyshev displays the most pronounced deviation from the standard scenario, underscoring its strong impact on non-linear structure formation.

Overall, these results indicate that halo formation at early times is enhanced in the CPL and Chebyshev cosmologies with respect to Planck-$\Lambda$CDM, while the $w$CDM model exhibits more moderate deviations, primarily affecting the low-mass end of the halo population. At later times, halo growth follows the standard hierarchical evolution common to all CDM-based cosmologies \citep{Mo:2010}, characterised by a progressive slowdown of individual accretion and an increasing relative contribution from merger-driven mass growth. Within this framework, the systematically higher abundance of massive haloes found in the CPL and Chebyshev models with respect to $\Lambda$CDM may indicate a more efficient late-time build-up of massive systems, with the effect being most pronounced in the Chebyshev case.

\subsubsection{Halo density profiles}

\begin{figure*}
  \centering
  \includegraphics[width=1.0\textwidth]{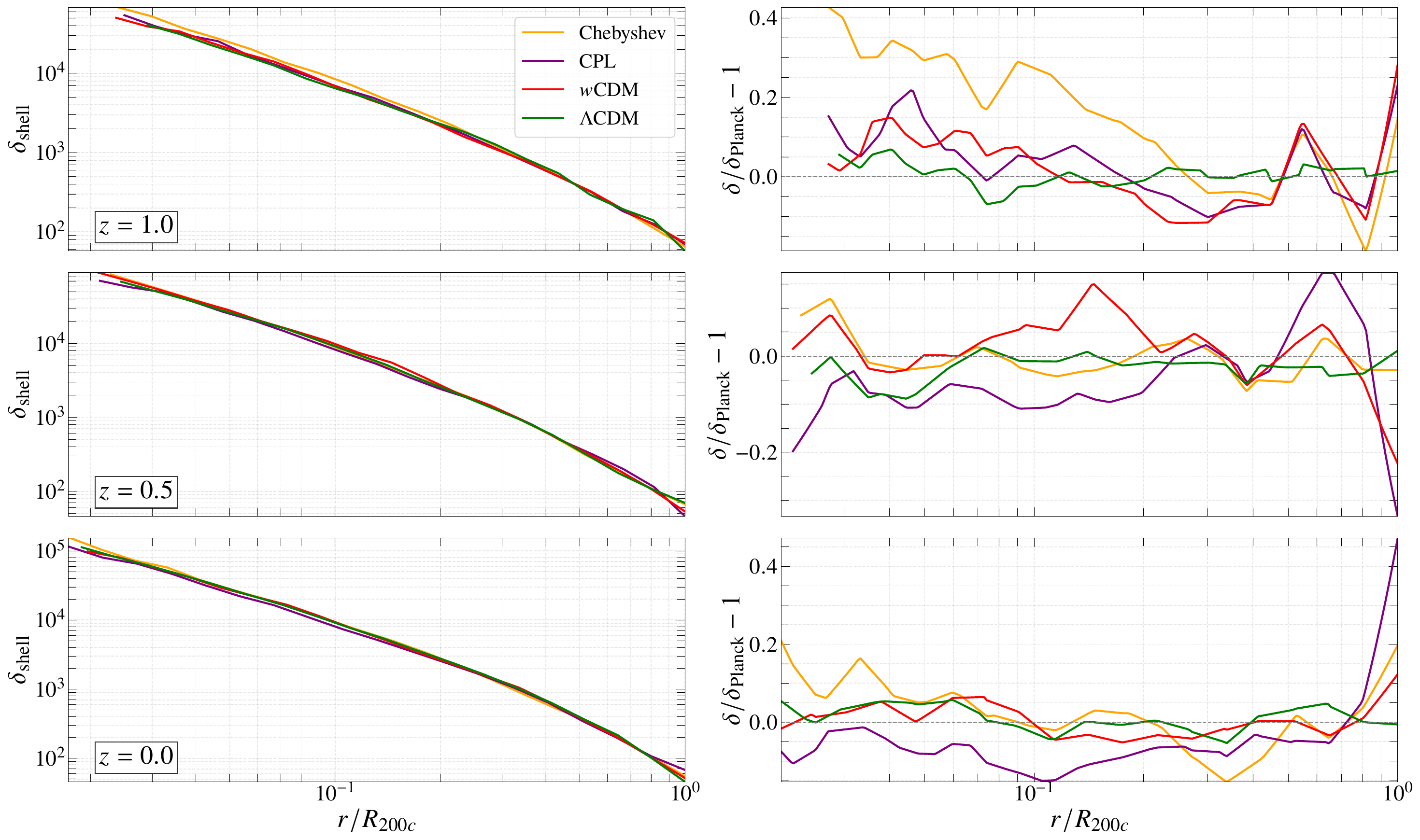}
  \caption{Shell density profiles, $\delta_{\rm shell} \equiv \rho_{\rm shell}/\rho_b$, as a function of the normalised radius $r/R_{200c}$ for tested cosmological models at six redshifts ($z=0.0,\;0.5,\;$ and $1.0$). {\it Left panels:} absolute shell overdensity. {\it Right panels:} relative difference of each model’s $\delta_{\rm shell}$ to that of Planck‑$\Lambda$CDM.}
  \label{fig:shell_profiles_most_massive}
\end{figure*}
\begin{figure*}
    \centering
    \includegraphics[width=1.\linewidth]{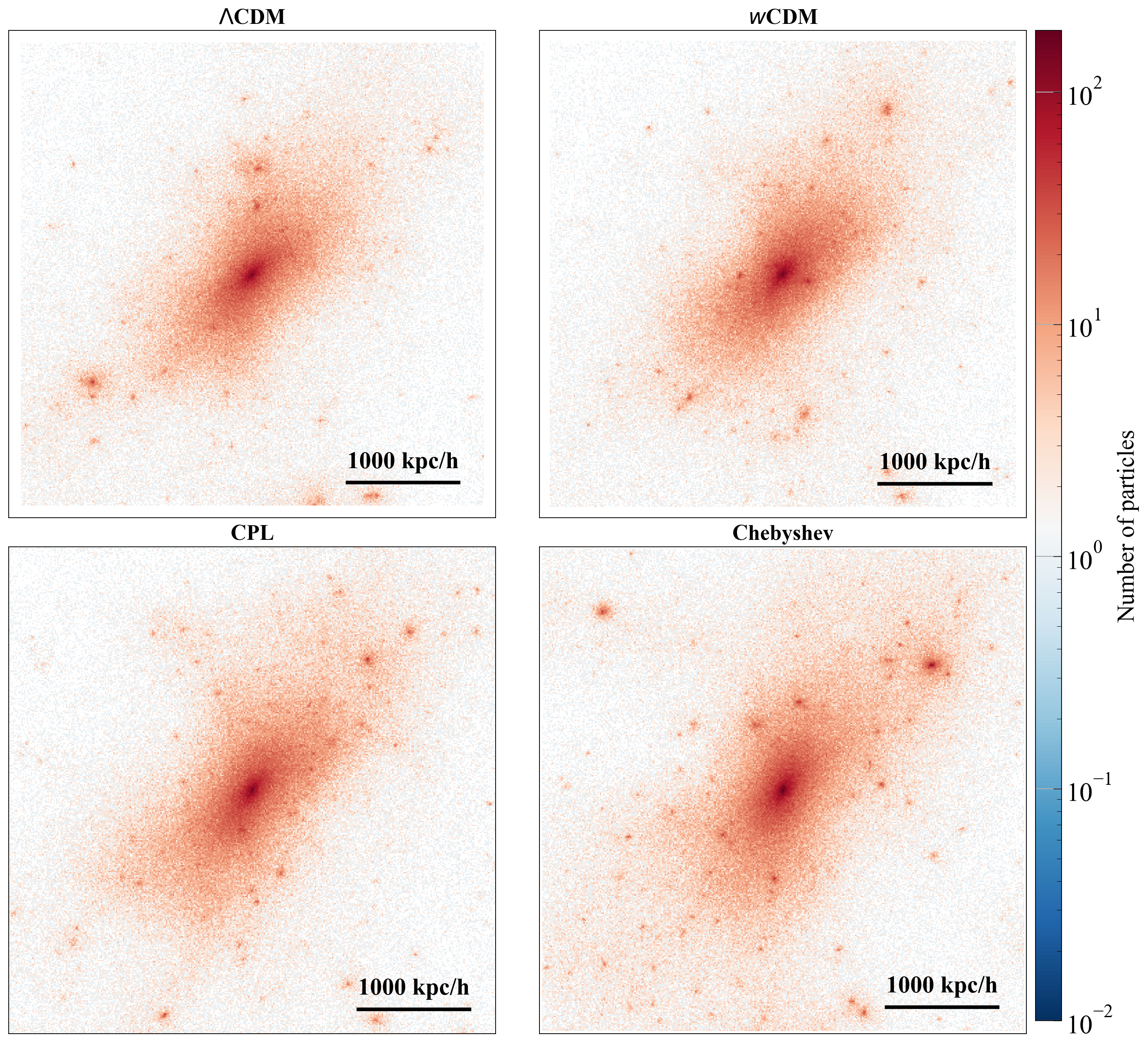}
    \caption{Particle number field of the most massive haloes of $\Lambda$CDM, $w$CDM, CPL and Chebyshev at $z=0$.}
    \label{fig:visual_most_masive_haloes}
\end{figure*}
\begin{figure*}
    \centering
    \includegraphics[width=.8\textwidth]{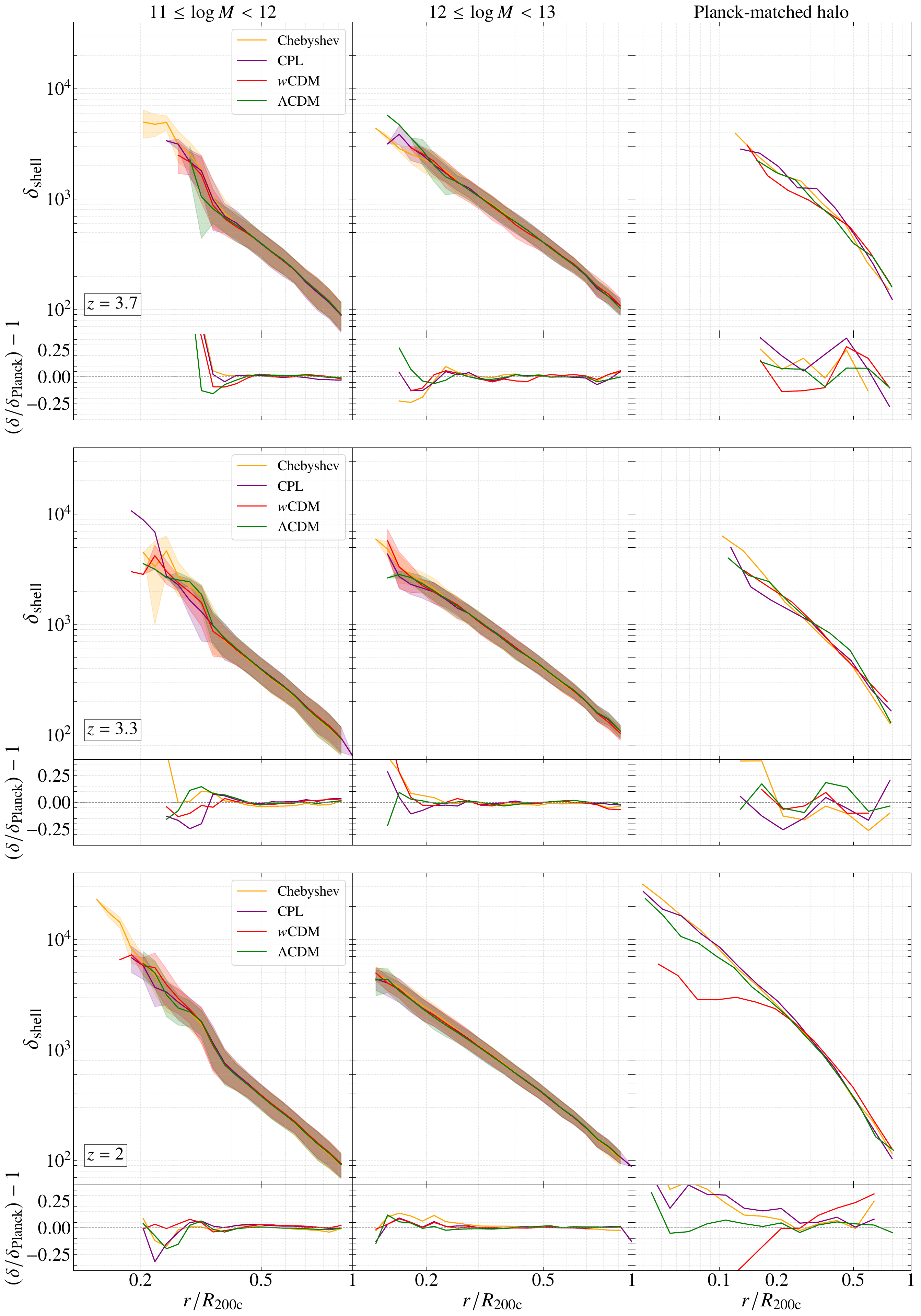}
    \caption{Shell overdensity profiles for dark energy haloes at high redshifts $z=3.7$, $3.3$ and $2.0$ (from top to bottom). In each row halo samples are divided into mass intervals; the left and middle columns show two mass bins ($11<\log M<12$, middle: $12<\log M<13$, with $M \equiv M_{\rm 200c}[M_\odot/h]$), while the right column displays only the most massive haloes (limited statistics at these redshifts). Upper panels present the mean overdensity profiles with the $1\sigma$ band and lower panels show the relative deviation with respect to the Planck--$\Lambda$CDM baseline. A sample of at least $\sim200$ haloes per mass range was considered.}
    \label{fig:dens_profiles_mean_highz}
\end{figure*}
\begin{figure*}
    \centering
    \includegraphics[width=.8\textwidth]{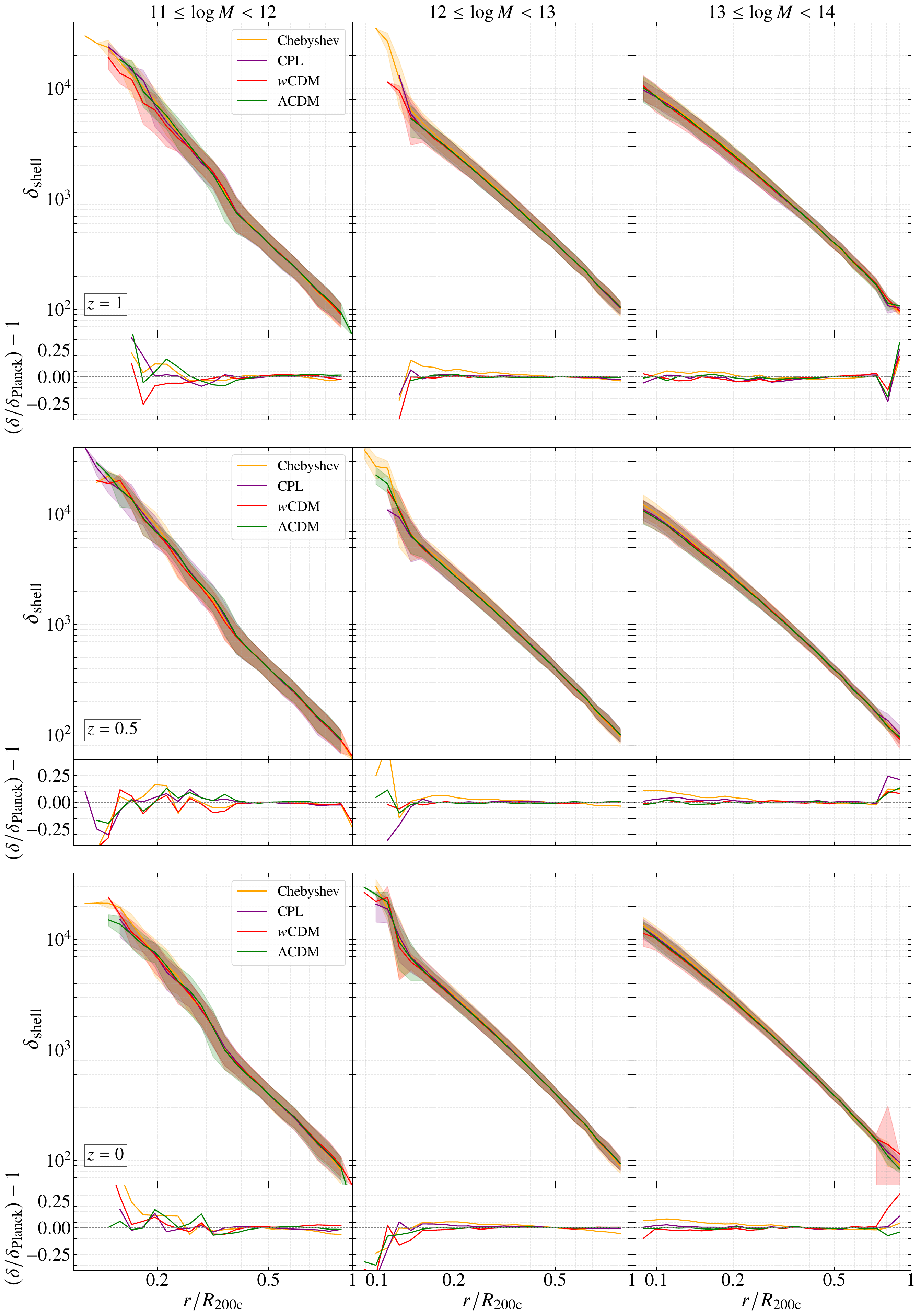}
    \caption{Same layout as Fig.~\ref{fig:dens_profiles_mean_highz} but for low redshifts $z=1.0$, $0.5$ and $0.0$ (from top to bottom). Unlike the high-redshift figure, here all three mass ranges are shown across the three columns (left: $11<\log M<12$, middle: $12<\log M<13$, right: $13<\log M<14$, with $M \equiv M_{\rm 200c}[M_\odot/h]$). Upper panels present the mean overdensity profiles with the $1\sigma$ band and lower panels show the relative deviation with respect to the Planck--$\Lambda$CDM baseline. A sample of at least $\sim200$ haloes per mass range was considered.}
    \label{fig:dens_profiles_mean_lowz}
\end{figure*}

We now present a comparison of the density profiles for all the models. 
We calculate the shell overdensity, defined as 
$\delta_{\rm shell} \equiv \rho_{\rm shell}/\rho_b$, 
in a shell of radius $r$, thickness $\Delta r$ and shell mass $M_{\rm shell}(r)\equiv M(r+\Delta r) -M(r)$,  which can be written as
\begin{equation}
    \delta_{\rm shell} = \frac{M_{\rm shell}(r)}{4\pi r^{2}\Delta r \, \rho_b},
\end{equation}
with $\rho_b = \Omega_{0m}\rho_{0c}(1+z)^3$ being the background density and $\rho_{0c}$ is the critical density at the present time.
First, in Fig.~\ref{fig:shell_profiles_most_massive} we present the individual overdensity profiles of the most massive haloes at low redshift. Also, we show a visual inspection of these haloes in Figure \ref{fig:visual_most_masive_haloes}. Then, we discuss the statistics of the overdensity profile (mean and standard deviation) at six redshifts (divided in two figures) 
for three mass intervals mentioned below\footnote{Note that Fig.~\ref{fig:dens_profiles_mean_highz} only shows the most massive halo instead a mass interval due to the lack of haloes in this mass interval.}. 
Figure \ref{fig:dens_profiles_mean_highz}, shows redshifts $z= 3.7,\; 3.3$ and $2.0$ and Figure \ref{fig:dens_profiles_mean_lowz}, shows redshifts $z=1.0, \;0.5\;$ and $0.0$, from the top to the bottom panels in both figures, respectively.

 Before discussing these results, it is important to note that \textsc{AHF} does not provide reliable measurements in the innermost regions of haloes. At these scales, the dynamics are dominated by two-body relaxation effects and the structure is therefore not converged according to the criterion of \citet{power:2003}. We define the converged region of each halo as the radial range $r > r_{\rm conv}$, where $r_{\rm conv}$ is the convergence radius determined by this criterion. In practice, $r_{\rm conv}$ is computed on a halo-by-halo basis using the convergence estimator implemented in \textsc{AHF} and all profile bins at radii $r < r_{\rm conv}$ are excluded from the analysis.

Fig.~\ref{fig:shell_profiles_most_massive} shows the shell overdensity profiles of the same most-massive halo evolved under the four different cosmological models. We use the same colour coding as in previous figures: green ($\Lambda$CDM), red ($w$CDM), purple (CPL) and orange (Chebyshev). Each row corresponds to a different redshift, $z=1.0$, $z=0.5$ and $z=0.0$, from top to bottom. The left column shows the absolute shell overdensity in logarithmic scale as a function of radius, $r\,[{\rm kpc}/h]$, while the right column presents the relative difference with respect to the Planck--$\Lambda$CDM reference, $\delta/\delta_{\rm Planck}-1$.

Halo matching is performed by starting from the most massive halo in the Planck--$\Lambda$CDM simulation and searching for spatially coincident counterparts in the alternative cosmological models using a position-based criterion. The search proceeds to lower-mass candidates only if needed and a match is accepted when a clear separation between the best and second-best candidates is satisfied. At low redshift, this procedure consistently selects the most massive halo in all cosmologies, confirming that massive haloes can be robustly matched across models for $z \le 1$.


For $z = 1$, Fig.~\ref{fig:shell_profiles_most_massive} shows an enhancement of the inner density profile of the most massive halo relative to the Planck reference in all extended models. The Chebyshev model exhibits the strongest and most radially extended enhancement, reaching $\sim 40\%$ across the inner region. In contrast, the CPL and $w$CDM models show more localised enhancements, confined to the smallest radii, at the $\sim 20\%$ and $\sim 15$--$20\%$ level, respectively. The $\Lambda$CDM profile remains nearly unchanged.

At lower redshifts ($z = 0.5$ and $z = 0$), the relative differences in the right column of Fig.~\ref{fig:shell_profiles_most_massive} are smaller than at $z = 1$ and broadly similar across models. The most systematic feature is a suppression of the CPL profile relative to the Planck reference in the inner regions, reaching a $\sim 20\%$ in the innermost point. The Chebyshev, $w$CDM and $\Lambda$CDM models remain broadly consistent with Planck within the level of fluctuations.

This assessment is based on a single halo and should therefore be regarded as indicative rather than statistical. The inferred ordering is consistent with the relative differences observed in the matter power spectrum and halo mass function. However, since the profiles are expressed in units of $r/R_{200c}$, part of the apparent inner enhancement may arise from the model dependence of $R_{200c}$ through the critical density $\rho_c(z)\propto H^2(z)$. In models where $H(z)$ is suppressed with respect to Planck, such as Chebyshev (see Fig.~\ref{fig:plots_background}), this leads to a larger $R_{200c}$, with $R_{200c} \propto H^{-2/3}(z)$, which can enhance the inner profile when expressed in scaled units.

To characterise halo behaviour across the full catalogue,
Figs.\ref{fig:dens_profiles_mean_highz} and
\ref{fig:dens_profiles_mean_lowz} present the shell overdensity
profiles at three high redshifts ($z=3.7$, $3.3$, $2.0$) and three
lower redshifts ($z=1.0$, $0.5$, $0.0$), respectively, for three halo
mass ranges. At each redshift, haloes are divided into three mass
intervals, $11\leq\log M_{\rm 200c}<12$, $12\leq\log M_{\rm 200c}<13$,
and $13\leq\log M_{\rm 200c}<14$, hereafter referred to as the low-,
intermediate- and high-mass bins. The upper subpanels in each row
show the mean overdensity profiles in each mass bin, with the shaded
region indicating one standard deviation $(\sigma)$, while the lower
subpanels display the relative deviation with respect to the Planck
baseline. Each mass interval contains at least
$\sim200$ haloes.
We note that, at high redshift ($2.0 \leq z \leq 3.7$), however, the number of
haloes in the highest-mass bin is insufficient for a statistically
robust stacking analysis. For this reason, the right column of
Fig.~\ref{fig:dens_profiles_mean_highz} shows instead a single
\textit{Planck-matched halo}, corresponding to the best-matched
counterpart across cosmologies.

We first examine the stacked overdensity profiles shown in the
left and middle columns of Fig.~\ref{fig:dens_profiles_mean_highz},
corresponding to the low- and intermediate-mass bins
($11 \le \log M < 12$ and $12 \le \log M < 13$). When expressed in
terms of the scaled radius $r/R_{200c}$, the stacked profiles display
a high degree of similarity across all cosmological models at the
three redshifts considered. In the low-mass bin
($11 \le \log M < 12$), the profiles are particularly consistent
across models over the radial range
$0.4 \lesssim r/R_{200c} \lesssim 0.7$, while for the
intermediate-mass bin ($12 \le \log M < 13$) this behaviour extends
over $0.2 \lesssim r/R_{200c} \lesssim 0.7$. Within these radial
intervals, deviations from the Planck--$\Lambda$CDM reference are
typically at the few-percent level and remain well within the
one-sigma scatter, indicating that the overall shape of halo density
profiles is largely insensitive to the underlying dark-energy model
once differences in halo size are accounted for.

In the low-mass bin ($11 \le \log M < 12$) of
Fig.~\ref{fig:dens_profiles_mean_highz}, the largest 
deviations are confined to the innermost radial bins
($r/R_{200c} \lesssim 0.4$). At $z=3.7$, the Chebyshev model tends to
reach slightly higher central overdensities than the other
cosmologies. However, this radial regime is also the most affected by
limited particle sampling, centring uncertainties and numerical
convergence constraints, particularly for low-mass haloes at high
redshift. As a result, the one-sigma scatter bands largely overlap
across all models and the apparent inner differences should be
interpreted with caution. At most, these trends are consistent with
modest variations in the inner structure of low-mass haloes rather
than with robust or systematic differences in their density profiles.

In the intermediate-mass bin ($12 \le \log M < 13$),
residuals relative to the Planck--$\Lambda$CDM reference remain small
at $z=3.7$ and decrease further at $z=3.3$ and $z=2.0$, reflecting a
progressive convergence of the profiles once scaled by $R_{200c}$,
particularly at intermediate and outer radii. A mild enhancement
persists for the Chebyshev model at small radii
($r/R_{200c} \lesssim 0.2$) at $z=3.3$ and $z=2.0$.

In the right column of Fig.~\ref{fig:dens_profiles_mean_highz},
we show a single Planck-matched halo per cosmological model.
Although potential halo progenitors can be independently identified
at high redshift in each simulation using AHF, we adopt here the same
position-based matching criterion used at low redshift ($z \le 1$) in
Fig.~\ref{fig:shell_profiles_most_massive} to track spatially
consistent counterparts across cosmologies at higher redshifts
($2 \le z \le 3.7$). While such counterparts can still be identified,
they no longer correspond to the most massive objects in the
simulations, since the correspondence with the most massive
low-redshift haloes can be reliably maintained only up to $z = 2$.

Within the Planck-matched halo column, the density profiles at
high redshift do not exhibit a common or monotonic radial trend
across cosmological models, as expected for a single matched system.
At $z=3.7$ and $z=3.3$, the profiles of all models display
fluctuations relative to the Planck--$\Lambda$CDM reference without a
clear systematic radial dependence. By contrast, at $z=2.0$, where
the Planck-matched haloes correspond to the most massive objects in
all cosmologies, the profiles become smoother and the comparison more
direct. In this case, the Chebyshev and CPL haloes exhibit enhanced
central densities relative to the reference, whereas the $w$CDM halo
exhibits a less dense core and comparatively higher densities at
larger radii.


We now consider the stacked overdensity profiles at lower redshifts, shown in Fig.~\ref{fig:dens_profiles_mean_lowz}, for $ z=1.0$, $ z=0.5$ and $z=0$. Compared to the high-redshift case, the profiles already exhibit significantly greater convergence in all three mass bins once radii are scaled by $R_{200c}$.

In the low-mass bin ($11 \le \log M < 12$), the stacked profiles are largely converged at the three redshifts. Residual differences are confined to the innermost resolved region ($r/R_{200c} \lesssim 0.4$), while the profiles remain consistent with the Planck--$\Lambda$CDM reference at intermediate and outer radii. These central features are characterised by substantial scatter and remain within the one-sigma bands, indicating that they may be dominated by numerical noise and halo-to-halo variations rather than by systematic effects in the structure formation process.

In the intermediate-mass bin ($12 \le \log M < 13$), small residual differences persist in the inner regions ($r/R_{200c} \lesssim 0.2$). The Chebyshev model shows a mild enhancement at the few-percent level at $z=1.0$, which remains visible down to $z=0.0$ over a limited radial range ($0.15 \lesssim r/R_{200c} \lesssim 0.3$). The remaining cosmological models exhibit a high degree of convergence over most radii, with only noisy and non-systematic fluctuations in the innermost bins.

For the high-mass bin ($13 \le \log M < 14$), the strongest degree of universality is observed at the three redshifts. The density profiles of all models are nearly indistinguishable over the full radial range. A modest central enhancement becomes slightly more noticeable at $z=0.5$ and $z=0.0$ for $r/R_{200c} \lesssim 0.3$, but remains at the few-percent level and within the statistical scatter.

Overall, the low-redshift profiles display a clear convergence toward a universal density structure across all cosmological models once radii are scaled by $R_{200c}$. Relative to the high-redshift results, deviations at low redshift are reduced in amplitude, increasingly confined to the innermost regions and progressively suppressed with increasing halo mass.


This behaviour is consistent with the well-established universality of halo density profiles in dissipationless hierarchical clustering within standard General Relativity \citep{nfw1997}. While the background expansion history affects the collapse redshift and therefore the halo concentration, the functional form of the virialised density profile is largely insensitive to moderate changes in cosmological parameters, reflecting the dominance of non-linear gravitational dynamics during halo assembly.

%

%

\section{Discussion and Conclusions}\label{sec:discussion}

In this study, we assessed the impact of four cosmologies ($\Lambda$CDM, $w$CDM, CPL and a Chebyshev expansion) on large-scale structure by (i) constraining background parameters with a joint analysis of BAO, CMB, CC and SLS observations and (ii) evolving those best-fitting cosmologies with numerical simulations to quantify differences in the dark matter field, the matter power spectrum, the halo abundances and overdensity profiles.

Using a joint Bayesian analysis, we have derived posterior constraints on a set of baseline $\Lambda$CDM parameters ($\Omega_{0b}, \Omega_{0\mathrm{CDM}}, h$) and on the additional degrees of freedom introduced in extended dark energy parametrisations (constant-$w$, CPL and a Chebyshev expansion). Our results confirm that the $\Lambda$CDM model remains broadly consistent with current background probes, but extended models permit modest, data-driven departures from the standard picture. The constant-$w$ model yields the largest statistically notable shift in the dark energy sector, with $w_0=-1.224^{+0.092}_{-0.100}$, a $\sim2.2\sigma$ departure from a cosmological constant and an associated upward shift in $h$ of $\sim1.6\sigma$. The CPL parametrisation remains statistically consistent with $\Lambda$CDM, primarily displaying enlarged parameter uncertainties rather than significant shifts. The Chebyshev expansion shows a notable offset in the baryon density (difference $\sim2.2\sigma$ $\Lambda$CDM) and a moderate deviation in the leading coefficient $C_0$ ($\sim1.16\sigma$), with higher-order coefficients consistent with $\Lambda$CDM within $1\sigma$. Cold dark matter and radiation densities remain largely compatible with the reference model across the board (all radiation deviations $\lesssim 0.4\sigma$).

The LSS simulations and analysis, on the other hand, illustrate the impact of density perturbations in the non-linear regime. A visual inspection of the dark matter field at $z=0$ reveals only subtle differences among models, except for the Chebyshev case, which exhibits the strongest contrast and more pronounced structures.

Following with the rest of the observables, we can conclude:

\begin{itemize}


\item The matter power spectrum reveals a clear hierarchy in the amplitude of density fluctuations across cosmological models. Although all scenarios share the same primordial power spectrum parameters, $(A_s,n_s)$, differences in the physical matter density $\Omega_{0m} h^2$ modify the transfer function and shift the epoch of matter--radiation equality, leading to distinct linear amplitudes already at high redshift. In models where $\Omega_{0m} h^2$ is larger than in the Planck--$\Lambda$CDM reference, matter--radiation equality occurs earlier, reducing the small-scale suppression imprinted in the transfer function and enhancing the linear power spectrum amplitude. 
In particular, the $w$CDM model exhibits enhanced power relative to the Planck--$\Lambda$CDM reference despite its lower matter density parameter $\Omega_{0m}$, owing primarily to its larger value of $\Omega_{0m} h^2$. For the CPL and Chebyshev parametrisations, additional modifications to $H(z)$ further modify the growth history, amplifying the cumulative differences at late times. The combined effect of early-time transfer-function differences and late-time growth modifications establishes the hierarchy in $\sigma_8$ observed at $z=0$, with Chebyshev showing the largest deviation from the Planck reference, followed by CPL and $w$CDM, while $\Lambda$CDM remains consistent with Planck.
    
\item The halo mass function reflects this hierarchy, but in a non-trivial manner. At high redshift ($2 \le z \le 3.7$), CPL and Chebyshev enhance the abundance of low- and intermediate-mass haloes, signalling earlier structure formation. At lower redshift ($z \le 1$), the excess shifts towards higher masses, particularly in the Chebyshev model, which exhibits a substantially larger abundance of the most massive systems. In contrast, $w$CDM shows a mild enhancement at the low-mass end but a suppression at higher masses. This behaviour illustrates that halo abundance depends not only on the fluctuation amplitude but also on the subsequent growth history and on the adopted halo mass definition. In $w$CDM, the smaller matter density parameter $\Omega_{0m}$ and the higher expansion rate $H(z)$ reduce the late-time growth efficiency. At the same time, the larger critical density $\rho_c(z)\propto H^2(z)$ leads to smaller $R_{200c}$ and therefore smaller enclosed masses for a given physical halo.


\item The internal density profiles of haloes remain broadly similar across cosmologies, as expected for models that share the same underlying gravitational theory \citep{nfw1997}. The modest differences observed are consistent with variations in assembly history and concentration rather than with modifications to the gravitational dynamics itself. Although changes in the background expansion history and in the physical matter density alter the timing of structure formation, the internal structure of virialised haloes shows a strong degree of convergence once scaled by $R_{200c}$.

\end{itemize}

Overall, we find that the Chebyshev model leads to the largest modifications in the structure formation. This behaviour can be attributed to the flexible nature of the Chebyshev parametrisation (see Eq.~\ref{eq:w1}): the zeroth coefficient, \(C_0 = 1.809^{+0.673}_{-0.723}\), constitutes the dominant departure from a cosmological constant within the allowed parameter space and is thus the primary driver of the large effects observed in the matter power spectrum. However, the polynomial basis itself also plays a role. Through the mapping \(x(z) \in [-1,1]\), 
combinations of \(C_1, C_2, C_3\) can generate non-monotonic features in \(w(z)\) within \(0 \le z \le z_{\max}\), while the matching conditions at \(z_{\max}\) can further modulate the expansion history. Establishing any additional impact of the polynomial shape would require targeted tests (e.g., fixing \(C_0 = 1\) or varying one coefficient at a time), which we leave for future work.

To conclude, within the observationally constrained alternative models explored in this work, background-level variations encoded in $\Omega_{0m} h^{2}$ and in the dark energy equation of state shape the expansion history and regulate the efficiency and timing of structure formation. These differences propagate coherently into non-linear observables such as the matter power spectrum and halo mass function, while the internal density profiles remain largely preserved under the same gravitational dynamics. This demonstrates that even moderate departures at the background level can generate measurable non-linear signatures, highlighting the complementarity between background probes and large-scale structure observables in constraining extended dark energy models.



We note the limitations of the present numerical study: the relatively small simulation volumes, modest particle numbers and single realisation per cosmology limit statistical robustness and the direct applicability of our quantitative results to survey-scale predictions. While our simulations suffice to reveal qualitative, model-dependent non-linear signatures—particularly for the Chebyshev case—definitive observational comparisons will require larger boxes, higher resolution and ensembles of realisations to fully characterise cosmic variance and systematics. Future work will extend the simulation suite to survey-like volumes (e.g., LSST/DESI footprints), incorporate baryonic effects via halo occupation modelling and combine predictions with targeted observables such as cluster counts and galaxy clustering to tightly test the extended dark energy parametrisations studied here.

\section*{Acknowledgments}

The authors are grateful to the anonymous referee for their insightful and constructive comments, which significantly improved the presentation and strengthened the results presented in this work. GAP acknowledges the support of the Chilean National Agency for Research and Development (ANID) through the ANID PhD Scholarship (Grant No. 21252489) and also thanks the Institute for Computational Cosmology at Durham University for providing workspace during the completion of this manuscript. MHA acknowledges support from \textit{Estancias Posdoctorales SECIHTI}. CB-H is supported by ANID through grant FONDECYT/Postdoctorado No. 3230512.

The data underlying this article will be shared on reasonable request to the corresponding author.

\bibliographystyle{mnras}
\bibliography{librero0}

\appendix


\end{document}